\documentstyle[prd,aps,eqsecnum,twocolumn,floats]{revtex}

\def\beq{\begin{equation}}
\def\eeq{\end{equation}}
\def\beqa{\begin{eqnarray}}
\def\eeqa{\end{eqnarray}}
\def\bfig{\begin{figure}}
\def\efig{\end{figure}}

\input{psfig}

\begin{document}
\fnsymbol{footnote}
%\draft
\wideabs{

\title{The $r$-modes in accreting neutron stars with magneto-viscous boundary layers}

\author{Justin B. Kinney${}^{1}$ and Gregory Mendell${}^2$}
\address{${}^1$Cornell University, Ithaca, NY 14853 \break ${}^2$LIGO Hanford Observatory,
P.O. Box 159 S9-02, Richland, WA 99352}

%\date{January 18, 2001}
\date{\today}
\maketitle
\begin{abstract}

We explore the dynamics of the $r$-modes in accreting neutron stars in two ways.  First,
we explore how dissipation in the magneto-viscous boundary layer (MVBL) at the crust-core interface governs
the damping of $r$-mode perturbations in the fluid interior. Two models are considered: one assuming an ordinary-fluid interior,
the other taking the core to consist of superfluid neutrons, type II superconducting protons, and normal electrons.
We show, within our approximations,
that no solution to the magnetohydrodynamic equations
exists in the superfluid model when both the neutron
and proton vortices are pinned.
However, if just one species of vortex is pinned, we can find solutions.
When the neutron vortices are pinned and the proton vortices are unpinned
there is much more dissipation than in the ordinary-fluid model, unless the pinning is weak.
When the proton vortices are pinned and the neutron vortices are unpinned
the dissipation is comparable or slightly less than that for the ordinary-fluid model,
even when the pinning is strong.
We also find in the superfluid model that relatively weak radial magnetic fields $\sim 10^9 \, {\rm G} \, (10^{8} \, {\rm K} /T)^2 $ greatly
affect the MVBL, though the effects of mutual friction tend to counteract the magnetic effects.
Second, we evolve our two models in time, accounting for accretion, and explore how the magnetic field strength,
the $r$-mode saturation amplitude, and
the accretion rate affect the cyclic evolution of these stars. If the $r$-modes control the spin cycles of
accreting neutron stars we find that magnetic fields can affect the clustering of the spin frequencies of
low mass x-ray binaries (LMXBs) and the fraction of these that are currently emitting gravitational waves.

\end{abstract}
}

%%%%%%%%%%%%%%%%%%%%%%%%%%%%%%%%%%%%%%%%%%%%%%%%%%%%%%%%%%%%%%%%%%%%%%%%%%%%
%%%
\section{Introduction}
\label{sectionI}

The $r$-modes are oscillation modes that occur in rotating fluids due to the Coriolis effect.
Great interest in these modes was generated after Andersson \cite{andersson} and Friedman and Morsink \cite{fried-morsink} showed
that gravitational-radiation backreaction tends to drive these modes unstable at all angular velocities.
However, internal dissipation most likely completely suppresses this instability in all
stars except neutron stars \cite{andersson,fried-morsink,lom,aks,owen-etal,lmo,lreview}.
The $r$-mode instability in neutron stars is further complicated by the fact that the problem splits into two cases:
superfluid neutron stars, with interior temperatures below approximately $10^9 \, {\rm K}$, and within which the interior
of a neutron star is expected to contain regions of superfluid neutrons mixed
with lower concentrations of superconducting protons, normal electrons and other exotic particles,
and ordinary-fluid neutron stars with interior tempertures above $10^9 \, {\rm K}$.
Furthermore, in a rotating neutron star the superfluid neutrons will form a dense array of quantized vortices, and if an interior magnetic field exists
the superconducting protons (for type II superconductivity) will form a dense array of flux tubes (also called vortices
in this paper). (See \cite{bppNature69,baym-patheck,alparlangersauls,pines-alpar,sauls,epstein} for discussions of superfluidity in neutron stars.)
A third case, not considered in this paper, is that in which the nucleons dissolve into a soup
of up, down, and strange quarks, which results in a strange star, not a neutron star.

The regimes within which unstable $r$-modes can exist in neutron stars have been narrowed down
as theoretical understanding has improved.
First, Bildsten and Ushomirsky \cite{bu} showed that when a solid crust is present,
the shear dissipation in the viscous boundary layer (VBL)
that forms at the crust-core interface greatly suppresses the $r$-mode instability.
Neutron stars are expected to form a solid crust for $\rho \lesssim
1.5 \times 10^{14} \, {\rm g/cm}^3$ and for tempertures
below an approximate melting temperature of $T \cong 10^{10} \, {\rm K}$
\cite{douchinhaensel,haensel}.
This work was extended by Andersson {\it et al.} \cite{ajks}, Rieutord \cite{rieutord},
Levin and Ushomirsky \cite{levinushomirsky}, Lindblom, Owen, and Ushomirsky \cite{lou}, and Mendell \cite{mendell2001}.
Bildsten and Ushomirsky \cite{bu} also predicted that magnetic fields would be important in the VBL, for
a magnetic field strength $B$ and temperature $T$, when $B \ge 10^{11} \, {\rm G} \, (10^{8} \, {\rm K} /T)$.
Mendell \cite{mendell2001} confirmed this and showed that magnetic fields further suppress the instability in
ordinary-fluid neutron stars.  Second, Jones \cite{jones2001a,jones2001b} and Lindblom and Owen \cite{lohyperons} have
shown that hyperon bulk viscosity further suppresses the instability above a temperature of $10^{9} {\rm K}$,
while Haensel, Levenfish, and Yakovlev \cite{haenseletalhyp} have shown that superfluidity of the baryons tends to suppress
hyperon bulk viscosity below $10^{9} {\rm K}$.  In this paper we ignore hyperon bulk viscosity altogether.
Finally, Wu, Matzner, and Arras \cite{wumatarras} and
Arras {\it et al.} \cite{arrasetal} have shown that even if the $r$-modes
are driven unstable, the saturation amplitude is likely to be very small.

However, the $r$-modes still remain important for the same reasons that sparked the initial interest in them.
First, it is still possible that they
play a role in the spin-down of hot young pulsars \cite{olrmodes}.  Second, they could be reponsible for
the clustering of spin frequencies inferred from the observations of low mass x-ray binaries
(LMXBs) \cite{vanderklis,strohmayer,wagoner84,bildsten,levin,ajks,whl,heyl}.  (However, note that
some complications exist with the interpretation of the observations; see \cite{jmk} and references therein.)
Finally, even given the recent results,
it is still possible that gravitational-waves from unstable small amplitude $r$-modes could be detected by enhanced or advanced
narrow-banded gravitational-wave detectors. Theoretical understanding of the $r$-modes instability is
far from complete, and thus more work is needed to understand these issues.

The purpose of this paper is to explore how the magneto-viscous boundary layer (MVBL) that forms at the crust-core interface in the presense
of a radial magnetic field affects the dynamics of the $r$-modes in accreting neutron stars.
Thus, this paper focuses on neutron stars with a solid crust ($T \le10^{10} \, {\rm K}$), and we assume that viscous dissipation
in the MVBL is the dominant form of dissipation in these stars.
Two models are considered: one assuming an ordinary-fluid interior;
the other, which we refer to as the superfluid model,
taking the core to consist of superfluid neutrons, type II superconducting protons, and normal electrons.
First, we explore how dissipation in the MVBL at the crust-core interface governs
the damping of $r$-mode perturbations in the fluid interior.   This extends the
work of Mendell \cite{mendell2001} to the superfluid case.
(Also, a minor coding error
in Mendell \cite{mendell2001} caused the MVBL damping times to come out about 53\%
percent too small and the critical angular velocities to come out about 11\%
too large.  We give the corrected results in this paper.)
Second, we evolve the $r$-mode amplitude, the angular velocity, and the temperature of the two models in time,
accounting for accretion. We explore how the magnetic field strength, the $r$-mode saturation amplitude, and
the accretion rate affect the cyclic evolution of these stars.  This extends the work of
Levin \cite{levin}, Anderson, {\it et al.} \cite{ajks}, and Wagoner, Hennawi, and Liu \cite{whl}
(see also Heyl \cite{heyl}).
We use the equations in Wagoner, Hennawi, and Liu \cite{whl} to evolve the models.

We show, within our approximations,
that no solution to the magnetohydrodynamic (MHD) equations
exists in the superfluid model when both the neutron
and proton vortices are pinned.
However, if just one species of vortex is pinned, we can find solutions.
When the neutron vortices are pinned and the proton vortices are unpinned
there is much more dissipation than in the ordinary-fluid model, unless the pinning is weak.
When the proton vortices are pinned and the neutron vortices are unpinned
the dissipation is comparable or slightly less than that for the ordinary-fluid model,
even when the pinning is strong.
We also find for the superfluid model that relatively weak radial magnetic fields $\sim 10^9 \, {\rm G} \, (10^{8} \, {\rm K} /T)^2 $ greatly
affect the MVBL.  We find that magnetic fields tend to make the critical angular velocity for the onset of the
$r$-mode instability temperature independent and increase the dissipation rate,
as Mendell \cite{mendell2001} found in the ordinary-fluid case, though the effects of mutual friction tend to counteract the magnetic effects.
Even when the magnetic field is zero, the correct scaling of the boundary layer thickness
with the proton mass density is given here for the superfluid model for the first time.

When we evolve our two models in time we find that the critical angular velocity decreases
with temperature sufficiently for all reasonable magnetic fields to produce the thermal run-away
found by Levin \cite{levin}.
(This happens even though magnetic fields tend to flatten the critical angular velocity vs. temperature curves.
Wagoner, Hennawi, and Liu \cite{whl} have shown that if the critical angular velocity is (mostly) temperature independent then
the $r$-mode amplitude and temperature oscillate with a period
of hundreds to thousand of years, while the spin frequency of the star stays roughly constant.)
Even with thermal run-way, Levin \cite{levin} and Anderson, {\it et al.} \cite{ajks} showed that the $r$-mode instability can still
produce clustering of the spin frequencies in LMXBs. If true, we find that magnetic fields can have important effects on this clustering.
For radial fields of $B \gtrsim 10^{11} \, {\rm G}$ in the ordinary-fluid model or for $B \sim 10^9 \, {\rm G}$ in the superfluid model
the spin cycle of an LMXB becomes thinner, which would cause LMXBs with these fields to cluster into a narrower range of spin frequencies.
Cumming, Zweibel, and Bildsten \cite{czb} have shown that buried fields in the crusts of accreting neutron stars
must be less than $10^{11} \, {\rm G}$, while typical external fields in LMXBs are $10^{8-9} \, {\rm G}$.
Furthermore, the interior magnetic
field may be expelled during neutron star spindown or affected by accretion \cite{srinietal,ruderman1991,rudermanetal97,konarbhatt}.
However, we know that neutron stars typically start out with large fields $\sim 10^{12} \, {\rm G}$.  Since magnetic field
evolution is uncertain, at some point it may be natural for the interior field to spend time at
the values given here. If the magnetic field narrows the spin cycle of an LMXB, this increases
the fraction of time it spends spinning down and radiating
significant amounts of gravitational radiation.  This, in turn, has an effect on
the number of LMXBs that are currently radiating.  Lowering
the saturation amplitude or increasing the accretion rate produces the same effects, making it difficult to infer
the interior field from current observations.  However, the detection of gravitational waves from the $r$-modes of an accreting neutron
star would determine the saturation amplitude (if the distance to source is known).
In this case it might be possible to place limits on the interior magnetic fields of LMXBs
with known accretion rates based on the observed allowed range of spin frequencies.

The next section reviews the Newtonian MHD equations for superfluid neutron stars.
These reduce to the ordinary-fluid case when appropriate limits
are taken.  Sec.~\ref{sectionII} finds an approximate solution to the MHD equations valid in the MVBL.
Sec.~\ref{sectionIII} presents the results of calculations for the MVBL damping times
and the critical angular velocity, and discusses limits on the magnetic field and the MVBL length-scales.
The models are evolved and the spin cycles of LMXBs are studied in Sec.~\ref{sectionV}.
Conclusions are discussed and suggestions for future work are made in Sec.~\ref{sectionVI}.

\section{Superfluid MHD Equations}
\label{sectionIb}

The Newtonian MHD equations for a mixture of superfluid neutrons, type II superconducting protons, and normal electrons
have been derived by Mendell \cite{mendell98}.  The quantities in the equations represent smooth averages
over volumes containing
many vortices, based on the approach of Bekarevich and Khalatnikov \cite{bekarevich-khalatnikov1961} and as
extended to neutron stars by Mendell and Lindblom \cite{mendell-lindblom1991}, Mendell \cite{mendell1991,mendell1991b},
and Lindblom and Mendell \cite{mendell-lind}.  It is convenient to define the average and relative
velocities, $\vec{v}$ and $\vec{w}$ respectively, in terms of the mass densities
and velocities of the neutron and protons:
\beqa
\rho \vec{v} = \rho_n \vec{v}_n + \rho_p \vec{v}_p, \label{defineveqn}
\eeqa
\beqa
\vec{w} = \vec{v}_p - \vec{v}_n. \label{defineweqn}
\eeqa
For a star rotating uniformly in equilibrium with angular velocity $\vec{\Omega}$ ($\Omega = |\vec{\Omega}|$),
the equations in Mendell \cite{mendell98} for small Eulerian perturbations (prefixed with a $\delta$)
can be written in the corotating frame as
\beqa
\partial_t \delta \vec{v} && + 2\vec{\Omega} \times \delta \vec{v} = - \vec{\nabla} \delta U
+ {1 \over \rho^2} \left ( {\partial \rho \over \partial \beta} \right )_p
\delta \beta \vec{\nabla} p \nonumber \\
&& + {1 \over \rho}
\left ( {\delta \vec{J} \over c} \times \vec{B} \right )
+ {1 \over \rho}\left [ (\vec{\nabla} \times \delta \vec{\lambda}_p)
\times {e \over m_p c} \vec{B} \right ] \nonumber \\
&& + {1 \over \rho} \vec{\nabla} \cdot (2 \eta_e \delta \tensor{\sigma}_e), \label{veqn}
\eeqa
\beqa
\partial_t \delta \vec{w} && + 2 \gamma \vec{\Omega} \times \delta \vec{w} =
- \vec{\nabla} \delta \beta
+ {1 \over \rho_p}
\left ( {\delta \vec{J} \over c} \times \vec{B} \right ) \nonumber \\
&& + {1 \over \rho_p} \left [ (\vec{\nabla} \times \delta \vec{\lambda}_p)
\times {e \over m_p c} \vec{B} \right ] \nonumber \\
&& + {1 \over \rho_p} \vec{\nabla} \cdot (2 \eta_e \delta \tensor{\sigma}_e) - {\rho \over \rho_p} \delta \vec{F}_n, \label{weqn}
\eeqa
\beqa
\partial_t \delta \vec{B} = \vec{\nabla} \times (\delta \vec{v}_e \times \vec{B}). \label{Beqn}
\eeqa
For simplicity, terms of order $\rho_e/\rho_p$ have been ignored, and only the largest vortex force
(dependent on $\vec{\lambda}_p$)
has been retained.  Dissipative effects due to the shear of the electron fluid, $\tensor{\sigma}_e$, and
the mutual friction force caused by electron scattering off the neutron vortices, $\vec{F}_n$, are included.
The mutual friction
force is given by \cite{mendell1991b}
\beqa
\delta \vec{F}_n = 2 \Omega \gamma {\cal B}_n \left [ \delta \vec{w} - { \vec{\Omega} (\vec{\Omega} \cdot \delta \vec{w})
\over \Omega^2} \right ], \label{FMutFric}
\eeqa
where ${\cal B}_n$ is the mutual friction coefficient.
In Eqs.~(\ref{veqn}) and ~(\ref{weqn}) $\delta U$ and $\delta \beta$ are related to the perturbed pressure, gravitational potential,
and chemical potentials in the star but play no further role in this paper (see Lindblom and Mendell \cite{mendell-lind}).
In Eqs.~(\ref{weqn}) and (\ref{FMutFric}) $\gamma$ is a dimensionless factor defined in Lindblom and Mendell \cite{lm2000} that arises due to
the ``entrainment'' effect that occurs due to strong interactions between neutrons and protons
\cite{andreevbashkin,alparlangersauls,sauls}.
When entrainment occurs the mass currents of the neutrons
and protons are given in terms of the superfluid velocities by,
\beqa
\vec{M}_n = \rho_{nn}\vec{v}_n + \rho_{np}\vec{v}_p, \label{MassnEqn}
\eeqa
\beqa
\vec{M}_p = \rho_{np}\vec{v}_n + \rho_{pp}\vec{v}_p. \label{MasspEqn}
\eeqa
The coefficients, $\rho_{nn}, \rho_{np}, \rho_{pp}$ form what is called the mass density matrix.
These are related to the ordinary mass densities and the entrainment factor $\gamma$ by,
\beqa
\rho_{nn} = \rho_n - \rho_{np}, \label{rhonneqn}
\eeqa
\beqa
\rho_{pp} = \rho_p - \rho_{np}, \label{rhoppeqn}
\eeqa
and
\beqa
\rho_{np} = {\rho_n \rho_p (1 - \gamma) \over \rho}. \label{rhonpeqn}
\eeqa
Next, note that Easson \cite{easson} calculates that the ratio of the magnetic diffusion time-scale
to the viscous diffusion time-scale in a typical neutron  star is roughly $10^{14} (10^8 \, {\rm K}/T)^4$.
Thus, the electrical conductivity of the electrons in the core and the crust is approximated as
infinite; magnetic diffusion plays no role for the temperatures and time-scales of interest in this paper,
and is ignored in Eq.~(\ref{Beqn}).  (See Mendell \cite{mendell2001} for further disscusion of this issue.)
Finally, $e$ is the absolute value of the charge of the electron, $m_p$ is the mass of the proton, and $c$ is the speed
of light.

The MHD limit is valid for studies of oscillations with phase velocities much less than the speed of light, frequencies
much less than the plasma and cyclotron frequencies, and large conductivities.  Under these circumstances the above equations,
along with the mass conservation laws and equations of state, completely determine the dynamics of system.  All the other
vector fields of interest are determined in terms of $\delta \vec{v}$, $\delta \vec{w}$, and $\delta \vec{B}$.
Specifically, for phase velocities much less than the speed of light, the displacement
current can be ignored in Amperes law, and the current density is given by
\beqa
\delta \vec{J} = {c \over 4 \pi} \vec{\nabla} \times \delta \vec{B}.
\eeqa
For an infinitely conducting crust, there will also be a surface current density
at the crust-core interface, given by
\beqa
\delta \vec{I} = {c \over 4 \pi} \Bigl [ \delta \vec{B} \times \hat{r} \Bigr ]_{r  = R_c}, \label{surfI}
\eeqa
where $R_c$ is the radius of the core.
For frequencies much less than the plasma and cyclotron frequency, the perturbed charge density is
neglibible, and the perturbed electrical current density is so small that the electron velocity is
approximately $\delta \vec{M}_p/ \rho_p$ \cite{mendell1991,mendell98}.
Using Eqs.~(\ref{defineveqn})-(\ref{defineweqn}) and Eqs.~(\ref{MasspEqn})-(\ref{rhonpeqn}) this can be written as
\beqa
\delta \vec{v}_e = \delta \vec{v} + {\rho_n \over \rho} \gamma \delta \vec{w} . \label{electveqn}
\eeqa
Finally, when the above approximations hold and the conductivity is high, electrons
(being the least massive charge carrier)
respond to make the Lorentz force on them negligible, and the electric field is given by
\beqa
\delta \vec{E} = - {\delta \vec{v}_e \over c} \times \vec{B} - {\vec{v} \over c} \times \delta \vec{B}.
\eeqa

Turning to the vortex force in Eqs.~(\ref{veqn}) and (\ref{weqn}), this force is due to
the underlying array the proton vortices (much smaller forces due to neutron vortices are ignored
in this paper).  It is given in terms of the vector vortex ``chemical potential'' $\lambda^a$
(basically the energy per unit length needed to increase the number of proton vortices by one
in direction a).  Mendell \cite{mendell1991,mendell98} shows that this force is given by
\beqa
&& \left [ (\vec{\nabla} \times \delta \vec{\lambda}_p)
\times {e \over m_p c} \vec{B} \right ]  \cong
- \left ( {\delta \vec{J} \over c} \times \vec{B} \right ) \nonumber \\
&& \qquad + {2m_p \over h}\left \{ \left [ \vec{\nabla} \times \delta (\varepsilon_p {\vec{B} \over B}) \right ]
\times {e \over m_p c} \vec{B} \right \}. \label{lambdapeqn}
\eeqa
The first term on the right side of this equation is not an approximation.  It occurs in the exact form
of the proton vortex force, and it always cancels the Lorentz force ($\propto \delta \vec{J} \times \vec{B}$)
in Eqs.~(\ref{veqn}) and (\ref{weqn}).  For this reason, Alfv\'{e}n waves do not occur in an exotic
type II superconducting proton, normal electron plasma.  Instead, these
waves are replaced by cyclotron-vortex waves (see Mendell \cite{mendell98} and the next section).
In the second term on the right side of Eq.~(\ref{lambdapeqn}), $\varepsilon_p$
is the energy per unit length of a proton vortex.
Small corrections to this term have been ignored (See Mendell \cite{mendell98}).
Finally, throughout this paper note that $B = |\vec{B}|$.

\section{Approximate Magneto-viscous Boundary Layer Solutions}
\label{sectionII}

Approximate solutions to the superfluid MHD equations are
found in this section.
The ordinary-fluid limit of the solutions is also found.
The equilibrium magnetic field is assumed to be arbitrary (for now) except that it is static in the corotating
frame, and it is restricted such that no equilibrium electrical currents exist.
This implies the equilibrium structure of the star is unchanged
by the presence of the magnetic field.
To facilitate the manipulation of tensor quantites, a rotating
coordinate basis will be used, and indices will be raised and lowered using the flat-space metric
tensor in spherical coordinates.  In this basis, the equilibrium velocity is $v^a = \Omega \phi^a$.
Following the notation of previous studies, note
that Latin indices are space indices, except where it is understood that
$n$, $p$, $e$, $o$, and $c$ refer to neutrons, protons, electrons, ordinary, and the crust respectively.

Let $\delta v^a$, $\delta w^a$, and $\delta B^a$ be
the standard $r$-mode solution valid in
the bulk of the core where viscous
and magnetic forces are small compared to the Coriolis force.
(As explained at the end of the last section, note that the magnetic forces
are due entirely to the proton vortex force in the superfluid model.)
By definition, the words ``standard $r$-mode solution'' mean the $r$-mode solution
that would exist in the fluid core, if no solid crust were present.
For the purposes of this paper, these solutions are also taken to be
the lowest order form of the standard solution, when the solution is expanded
in powers of the angular velocity.
The standard $r$-mode solution is already known from
previous studies of the $r$-modes.
Note that $\delta B^a$ is approximately zero,
and is taken to be exactly zero in the standard $r$-mode solution.
Furthermore, Lindblom and Mendell \cite{lm2000} show that
$\delta w^a = 0$ at lowest order, and thus the lowest
order superfluid $r$-mode solution is identical to the
lowest order ordinary-fluid solution described in previous $r$-mode papers.
(In general, see reference \cite{lm2000} for the standard $r$-mode solution in the notation used in this paper.)
However, because of the solid crust,
boundary conditions must be applied to the tangential components of the velocities at the crust-core interface.
An ordinary viscous fluid cannot slip at a perfectly rigid boundary, and superfluid vortices
cannot move if perfectly pinned at the boundary.  However, electrons can slip at a conducting
boundary. When fluids or vortices cannot slip at a boundary this causes a boundary layer to form.
In the boundary layer the magnitudes of the viscous, magnetic, and Coriolis forces become comparible.
Mutual friction forces can be important too. Thus, all these forces have effects on the structure of the
boundary layer. When both viscous and magnetic forces exist we refer to the boundary layer as a MVBL.

Let $\delta \tilde{v}^a$, $\delta \tilde{w}^a$, and $\delta \tilde{B}^a$
be the corrections that must be added to the standard $r$-mode solution to enforce
the tangential boundary conditions at the crust-core interface.
The problem in this section reduces to finding
solutions for the corrective quantities,
$\delta \tilde{v}^a$, $\delta \tilde{w}^a$, and $\delta \tilde{B}^a$.
Since the total fields are
$\delta v^a + \delta \tilde{v}^a$, $\delta w^a + \delta \tilde{w}^a$,
and $\delta B^a + \delta \tilde{B}^a$, and the equations are linear, the corrective
quantities obey the same equations as the standard quantities,
Eqs.~(\ref{veqn})-(\ref{Beqn}).
However, approximations that hold true for the corrective quantities in the MVBL
are made when solving for these quantities that differ from the approximation made
when finding the standard quantities.
(This is done such that total solution is approximately valid inside and outside the MVBL.)
The main assumption, following the work of Lindblom, Owen, and Ushomirsky \cite{lou} and references therein,
is that the boundary conditions force the corrective quantities to rapidly change in the radial direction.
Thus, it is assumed that terms involving radial derivatives
of the corrective quantities dominate the angular derivatives of these quanties,
and dominate any derivative of all other quantities.
(For certain angles, certain caveats must be added to this statement.
These angles are discussed in Secs.~\ref{sectionIII}.)
By construction, the ratio of radial derivative terms kept
to the terms dropped will be the ratio of the core radius to the boundary-layer length-scales.
Thus, the main assumption is valid as long as the boundary layer length-scales are much less
that the core radius.

Thus, it is possible to find approximate equations for the corrective quantities,
$\delta \tilde{v}^a$, $\delta \tilde{w}^a$, and $\delta \tilde{B}^a$
by taking the following steps.
First, let all perturbed quantities in the corotating frame have
time dependence ${\rm exp}(i\kappa \Omega t)$, where $\kappa$ is a constant that gives
the mode frequency in terms of $\Omega$.
Second, note that
since in the standard solution $\delta v^r$, $\delta w^r$, and $\delta B^r$ vanish everywhere, mass is conserved at the boundary
and the divergence of the magnetic field is zero if the
radial components, $\delta \tilde{v}^r$, $\delta \tilde{w}^r$, and $\delta \tilde{B}^r$ also vanish.
Finally, of the spacial derivative terms, only the highest order radial derivitives acting
on the corrective quantities are kept, all other spacial derivative terms are dropped.
Taking these steps, the resulting equations are as follows.
First, Eq.~(\ref{Beqn}) becomes,
\beqa
\delta \tilde{B}^a = - {i \over \kappa \Omega} B^r\partial_r \delta \tilde{v}_e^a.
\eeqa
Substituting this, Eq.~(\ref{lambdapeqn}), and Eq.~(\ref{FMutFric}) into Eq.~(\ref{veqn}) and Eq.~(\ref{weqn}) gives
\beqa
&& \rho ( i\kappa\Omega \delta \tilde{v}^\theta - 2 \Omega {\rm cos}\theta {\rm sin}\theta \delta \tilde{v}^\phi )/\rho_p =
{\cal F}  \partial_r^2 \delta \tilde{v}_e^\theta , \label{Vthetaeqn}
\eeqa
\beqa
&& \rho ( i\kappa\Omega \delta \tilde{v}^\phi + 2 \Omega {\rm cot}\theta \delta \tilde{v}^\theta )/\rho_p =
{\cal F}  \partial_r^2 \delta \tilde{v}_e^\phi , \label{Vphieqn}
\eeqa
\beqa
&& i\kappa\Omega \delta \tilde{w}^\theta - 2 \gamma \Omega {\rm cos}\theta {\rm sin}\theta \delta \tilde{w}^\phi \nonumber \\
&& \qquad + \, (2 \gamma \Omega \rho {\cal B}_n / \rho_p) \delta w^\theta = {\cal F} \partial_r^2 \delta \tilde{v}_e^\theta
, \label{Wthetaeqn}
\eeqa
\beqa
&& i\kappa\Omega \delta \tilde{w}^\phi + 2 \gamma \Omega {\rm cot}\theta \delta \tilde{w}^\theta \nonumber \\
&& \qquad + \, (2 \gamma \Omega \rho {\cal B}_n / \rho_p) \delta w^\phi = {\cal F} \partial_r^2 \delta \tilde{v}_e^\phi
 , \label{Wphieqn}
\eeqa
where
\beqa
{\cal F} = -i\left [{V_{\rm CV}^2 \over \kappa \Omega} \left ( {B_r \over B} \right )^2
+ {i \eta_e  \over \rho_p} \right ] . \label{calFeqn}
\eeqa
In this equation, $V_{\rm CV}^2$ is the square of the cyclotron-vortex wave speed, defined by \cite{mendell98}
\beqa
V_{\rm CV}^2 \equiv {\varepsilon_p B \over \Phi_0 \rho_p}, \label{SecIIVCVeqn}
\eeqa
where the quantum of flux is $\Phi_0 \equiv hc/2e$, and $\eta_e$ is the electron
viscosity (and recall that $\varepsilon_p$ was previously defined after Eq.~[\ref{lambdapeqn}]).

Equating the right sides of Eqs.~(\ref{Vthetaeqn}) and (\ref{Wthetaeqn}), and
the right sides of Eqs.~(\ref{Vphieqn}) and (\ref{Wphieqn}),
the components $\delta \tilde{w}^\theta$ and $\delta \tilde{w}^\phi$ are given algebraically by
\beqa
&& i\kappa\Omega \rho_p \delta \tilde{w}^\theta - 2 \gamma \Omega \rho_p {\rm cos}\theta {\rm sin}\theta \delta \tilde{w}^\phi
+ 2 \gamma \Omega \rho {\cal B}_n \delta w^\theta \nonumber \\
&& \qquad = \rho ( i\kappa\Omega \delta \tilde{v}^\theta - 2 \Omega {\rm cos}\theta {\rm sin}\theta \delta \tilde{v}^\phi )
, \label{alegbraWthetaeqn}
\eeqa
and
\beqa
&& i\kappa\Omega \rho_p \delta \tilde{w}^\phi + 2 \gamma \Omega \rho_p {\rm cot}\theta \delta \tilde{w}^\theta
+ 2 \gamma \Omega \rho {\cal B}_n \delta w^\phi = \nonumber \\
&& \qquad \rho ( i\kappa\Omega \delta \tilde{v}^\phi
+ 2 \Omega {\rm cot}\theta \delta \tilde{v}^\theta )
. \label{algebraWphieqn}
\eeqa
Combining these with Eq.~(\ref{electveqn}) shows that Eqs.~(\ref{Vthetaeqn}) and Eqs.~(\ref{Vphieqn}) are a 4th order system for
$\delta \tilde{v}_e^\theta$ and $\delta \tilde{v}_e^\phi$.

In deriving the above equations a factor of ${\rm cos}^2 \theta$ has been ignored
in the mutual friction force term in Eq.~(\ref{Wthetaeqn}).  Also,
\beqa
\delta \left( {B^a \over B} \right ) = {\delta B^a \over B}
 - B^a { B^b \delta B_b \over B^3}, \label{Bprojected}
\eeqa
which appears in the proton vortex force, has been approximated as $\delta B^a/B$.
In the final equations this corresponds to ignoring
terms of the order $(B_r/B)^2(B^\theta/B^2)^2$, $(B_r/B)^2(B^\theta B^\phi/B^2)$, and $(B_r/B)^2(B^\phi/B)^2$.
These rather crude approximations make the analysis much easier
to understand. The equations and solutions without these approximation are given in the Appendix.
(In the final analysis, all magnetic terms, included the ignored ones, vanish if $B_r$ vanishes; $B_r$ controls
the magnetic effects on the MBVL.  Thus keeping only the largest terms that depends on $B_r$ is not too bad.)

The corrective boundary layer solution is then found by allowing all perturbative quantities
to vary as ${\rm exp}[ik(R_c - r)]$ (recall $R_c$ is the radius of the core).
It is easy to show that solutions exist for
\beqa
k_\pm = K_\pm\sqrt{ {\Omega
\over
{V_{\rm CV}^2 \over \kappa \Omega}\left ( {B_r \over B} \right )^2 + {i \eta_e \over \rho_p} } } , \label{kpmeqn}
\eeqa
\beqa
&& K_\pm = \sqrt{ {(\kappa \pm 2{\rm cos}\theta)[\kappa \pm 2 \gamma {\rm cos} \theta - (2 i \rho \gamma {\cal B}_n/\rho_p)]
\over [(\rho_n \gamma + \rho_p)/\rho]\kappa \pm 2 \gamma{\rm cos}\theta - 2 i \gamma {\cal B}_n} } , \label{Keqn}
\eeqa
\beqa
\delta \tilde{v}^\theta = \pm i {\rm sin}\theta \delta \tilde{v}^\phi,  \label{vthetaphieqn}
\eeqa
\beqa
\delta \tilde{w}^\theta= \pm i {\rm sin}\theta \delta \tilde{w}^\phi.  \label{wthetaphieqn}
\eeqa
and
\beqa
\delta \tilde{w}^\theta = { \rho( \kappa \pm 2 {\rm cos}\theta)
\over \rho_p(\kappa \pm 2 \gamma {\rm cos}\theta) - 2 i \rho \gamma {\cal B}_n}\delta \tilde{v}^\theta .
\eeqa
Choosing solutions where ${\rm Im}(k_\pm) > 0$, so that the solution
decays exponentially as $r \rightarrow 0$, the general solution of the equations is of the form
\beqa
\delta \tilde{v}_e^\theta = [ C_+ e^{ik_+ (R_c - r)}
+ C_- e^{ik_- (R_c - r)} ] e^{i \kappa \Omega t} .
\eeqa
We choose to define $C_\pm$ as the constants that give
$\delta \tilde{v}_e^\theta$.  Once these are determined by the tangential boundary conditions at the crust-core interface,
all other components and velocities are determined in terms of these constants.

It is easy to find the analogous solution for a mixture of ordinary-fluid neutrons,
protons, and electrons.  In this case, all the fluids act as a single fluid, and
the equations for this case can be obtained from
(\ref{Vthetaeqn}) and (\ref{Vphieqn}) by taking
the following limits:
\beqa
&& \lambda_p^a \rightarrow 0, \nonumber \\
&& \rho_p \rightarrow \rho,  \nonumber \\
&& \eta_e \rightarrow \eta, \nonumber \\
&& v_e^a \rightarrow v^a, \nonumber \\
&& V_{\rm CV}^2 \rightarrow V_{\rm A}^2,
\eeqa
where $V_{\rm A}^2$ is the square of the Alfv\'{e}n wave speed, defined as
\beqa
V_{\rm A}^2 \equiv {B^2 \over 4 \pi \rho}. \label{VAeqn}
\eeqa
This equation should be compared with Eq. (\ref{SecIIVCVeqn}) for $V_{\rm CV}^2$.
Choosing solutions where ${\rm Im}(k^o_\pm) > 0$, so that the solution
decays exponentially as $r \rightarrow 0$, the general solution for
the ordinary-fluid equations is of the form
\beqa
\delta \tilde{v}^\theta = [ C_+ e^{ik^o_+ (R_c -r)}
+ C_- e^{ik^o_- (R_c - r)} ] e^{i \kappa \Omega t} ,
\eeqa
where
\beqa
k^o_\pm = K^o_\pm \sqrt{ {\Omega
\over
{V_{\rm A}^2 \over \kappa \Omega}\left ( {B_r \over B} \right )^2 + {i \eta \over \rho}}} , \label{kopmeqn}
\eeqa
and
\beqa
&& K^o_\pm = \sqrt{\kappa \pm 2{\rm cos} \theta} . \label{Koeqn}
\eeqa
These equations are exactly those obtained by Mendell \cite{mendell2001}.

We must now consider what the correct tangential boundary conditions are at the crust-core interface.  In the ordinary fluid case
it is the no-slip boundary condition for a viscous fluid
\beqa
\left [ \delta v^a + \delta \tilde{v}^a = (1 - {\cal S}) \delta v^a \right ]_{r = R_c} . \label{ordnoslip}
\eeqa
Following Levin and Ushomirsky \cite{levinushomirsky}, but introducing our own notation, we introduce the
``slip factor'' ${\cal S}$ into Eq.~(\ref{ordnoslip}). This is a simplified way to account for the motion of the crust.
When ${\cal S} = 1$ the crust
is perfectly rigid, and no slipping of the fluids at the boundary is allowed. A value of ${\cal S} = 0$ would correspond to a fluid crust,
in which case the corrective solutions vanish.  However, a realistic crust is not a fluid and not perfectly rigid.
Levin and Ushomirsky \cite{levinushomirsky} found for a toy model that $0.05 \le {\cal S} \le 1$.

Note that in the superfluid case the simple no-slip boundary condition does not apply.  The reasons are the neutron and protons are superfluid and
thus can slip, and the electrons are conducting and thus can slip along the conducting crust too.
Instead, the tangential boundary conditions at the crust-core interface are determined by the pinning of the neutron
and proton vortices.  However, if both are pinned then the problem is over-determined and no solution exist.
The reason for this is that in our approximation scheme the only independent velocity is that of the electrons; the other
velocities are determined algebraically in terms of the components of $\delta \vec{v}_e$.  Thus, if both the neutron and protron vortices are pinned,
the approximation scheme used here breaks down and either the MVBL becomes much more complicated or the $r$-modes are drastically changed.

However, if just one species of vortex is pinned, we can find solutions.  (More realistically, such solutions probably
hold if the pinning of one vortex species completely dominates over the other species.)
Using the equations in the Appendix of Mendell \cite{mendell1991b} we find, for a particular species, that the
smooth-averaged perturbed vortex core velocity is proportional to the perturbed mass
current projected perpendicular to the equilibrium direction of the vortex array.  Since the radial components of the velocities are
zero, this amounts to applying the tangential boundary conditions to the mass currents themselves.
Using Eqs.~(\ref{MassnEqn}) and (\ref{MasspEqn}), pinning just one vortex species corresponds to the pinned-vortex boundary condition,
\beqa
\left [ \delta v^a + {\rho_{nn} \over \rho_n} \delta \tilde{v}_{n}^a  + {\rho_{np} \over \rho_n} \delta \tilde{v}_{p}^a
=  (1 - {\cal S}) \delta v^a \right ]_{r = R_c} , \label{supNnoslip}
\eeqa
for pinned neutron vortices,
and
\beqa
\left [ \delta v^a + {\rho_{np} \over \rho_p} \delta \tilde{v}_{n}^a  + {\rho_{pp} \over \rho_p} \delta \tilde{v}_{p}^a
=  (1 - {\cal S}) \delta v^a \right ]_{r = R_c}, \label{supPnoslip}
\eeqa
for pinned proton vortices.  We introduce the factor ${\cal S}$ again to account for slipping of vortices or motion of the crust.
A value of ${\cal S} = 1$ corresponds to strong (perfect) pinning while a value of ${\cal S} = 0$ corresponds to no pinning.  Small
values of ${\cal S}$ correspond to weak pinning. Several studies imply that vortex pinning in
neutron stars, in general, may be moderate or weak, \cite{pizzocheroetal,sedrakianetalpin,linkcutlerpin} though certain configurations
may result in increased pinning strength \cite{hirasawashibazaki}.
(Also see Ruderman, Zhu, and Chen \cite{rudermanetal97} for a disscussion of pinning and vortex interactions in the core.)

Applying the no-slip or pinned-vortex boundary conditions give
\beqa
C_\pm = -{1 \over 2}{\cal S}\Lambda_\pm(\delta v^\theta \pm i {\rm sin}\theta \delta v^\phi), \label{Cpmeqn}
\eeqa
where
\beqa
\Lambda_\pm = 1, \label{LambdaordPinP}
\eeqa
in the ordinary fluid case or when the proton vortices are pinned, and
\beqa
\Lambda_{\pm} = { [(\gamma\rho_n + \rho_p)/\rho] \kappa \pm 2\gamma \cos\theta - 2i\gamma {\cal B}_n \over (\rho_{np}/\rho_n)\kappa - 2i\gamma {\cal B}_n } , \label{LambdaordPinN}
\eeqa
when the neutron vortices are pinned. In this latter case, this factor represents the fact that since equal but opposite forces couple the neutrons to
the charged fluids (via the entrainment effect and mutual friction)
the accelerations and motions of these fluids are inversely proportional to their mass densities.  Thus,
the factor $\Lambda_\pm$ is quite large when the neutron vortices are pinned.
In contrast, since the electrons move with the proton mass currents, $\Lambda_\pm$ is
$1$ when the proton vortices are pinned.  Similarly, since all the fluids move together in the ordinary-fluid case $\Lambda_\pm$ is $1$ in this case as well.
Furthermore, note that $\delta v^\theta$ and $\delta v^\phi$ in Eq.~(\ref{Cpmeqn})
are the same for the superfluid and ordinary-fluid models because, as already pointed out,
the lowest-order standard $r$-mode solutions are identical for these two cases.
These components of the velocity
in the corotating frame (with their time dependence cancelled) are
\beqa
\delta v^\theta = -i A r^{m - 1} {\rm sin}^{m-1}\theta e^{im \phi}, \label{Vrmodetheta}
\eeqa
\beqa
{\rm sin}\theta \delta v^\phi = A r^{m - 1} {\rm sin}^{m-1}\theta {\rm cos} \theta e^{im \phi} . \label{Vrmodephi}
\eeqa

As shown in previous papers, the viscous damping rate is given by integrating the shear over the core:
\beqa
{1 \over \tau_v} = {1 \over 2 E}\int 2 \eta \delta \sigma^*_{ab} \delta \sigma^{ab} d^3x, \label{oneovertauv}
\eeqa
where $E$ is the energy of the mode as defined in e.g., Lindblom and Mendell \cite{lm2000}, but limited to the core.
The largest contribution to the integral comes from the radial derivatives of
the corrective boundary layer velocities, so that in the superfluid case
\beqa
\delta \sigma^*_{ab} \delta \sigma^{ab} = {1 \over 2} R_c^2 (|\partial_r \delta \tilde{v}_e^\theta|^2
+ |\partial_r {\rm sin}\theta \delta \tilde{v}_e^\phi|^2), \label{shearsquared}
\eeqa
ignoring terms smaller than these by a factor of the boundary layer length-scales over the core radius.
Thus, it can be shown that the solutions presented in this section give a VBL damping time of
\beqa
\tau_v = {2 \pi \over \eta_e {\cal S}^2 {\cal I}} {2^{m + 3} (m + 1)! \over m (2m + 1)!!}
\int_0^{R_c} \rho \left ( {r \over R_c} \right )^{2m + 2} dr,
\label{tauv}
\eeqa
where
\beqa
&& {\cal I} = \int_0^{2\pi} \int_0^\pi \Bigr [ |\Lambda_+|^2 |k_+|^2 d_+ (1 - {\rm cos}\theta)^2 \nonumber \\
&& \qquad + \, |\Lambda_-|^2 |k_-|^2 d_- (1 + {\rm cos}\theta)^2 \Bigl ] {\rm sin}^{2m -1} \theta d\theta d\phi.
\label{Im}
\eeqa
Using the definition
\beqa
k_\pm \equiv {2 \pi \over \lambda_\pm} + { i \over d_\pm}, \label{lambdaandddefined}
\eeqa
the boundary layer thicknesses in Eq.~(\ref{Im}) and boundary layer wavelengths are defined as
\beqa
\lambda_\pm \equiv {2 \pi \over {\rm Re}(k_\pm)}, \label{lamdefined}
\eeqa
\beqa
d_\pm \equiv {1 \over {\rm Im}(k_\pm)}. \label{ddefined}
\eeqa
The results given here reduce to those given in Mendell \cite{mendell2001} in the ordinary-fluid limit.

Before we study in detail how MVBL damping affects the $r$-modes, we note the following
useful analytic expressions for $\tau_v$ when the mutual friction coefficient ${\cal B}_n$ is $0$.
For simplicity we choose $B = B_r =$ constant. This is unrealistic (and unphysical
unless as many field lines enter the crust core interface as leave it) but Mendell \cite{mendell2001} has shown that this simple model gives
the same qualitative results as a more complicated dipole field.  In this case, the factors next to $K_{\pm}$ in Eq.~(\ref{kpmeqn})
and next to $K_{\pm}^o$ in Eq.~(\ref{kopmeqn}) are independent of $\theta$, and we can factor them out of the integral ${\cal I}$, giving us
\beqa
&& \frac{1}{\tau_v} = {\cal S}^2 \Biggr [ C_{OF} {\rm Re}\left(\sqrt{q_{OF} B_r^2 + i \Omega/T_{10}^2}\right) \nonumber \\
&& \qquad + D_{OF} {\rm Im}\left(\sqrt{q_{OF} B^2 + i \Omega/T_{10}^2}\right) \Biggl ], \label{OFtau_v}
\eeqa
for the ordinary-fluid model and
\beqa
&& \frac{1}{\tau_v} = {\cal S}^2 \Biggr [ C_{SF} {\rm Re}\left(\sqrt{q_{SF} B + i \Omega/T_8^2}\right) \nonumber \\
&& \qquad + D_{SF} {\rm Im}\left(\sqrt{q_{SF} B + i \Omega/T_8^2}\right) \Biggl ] , \label{SFtau_vnoMF}
\eeqa
for the superfluid model.
In these equations $q_{OF} \equiv 1/(4\pi\kappa\eta_0)$, and $q_{SF} \equiv \epsilon_p/(\Phi_0\kappa\eta_{e,0})$, where $\eta_0$ and $\eta_{e,0}$
are the temperature independent part of the viscosities such that $\eta = \eta_{0} (10^{10} \, {\rm  K} / T)^{2}$ and
$\eta_{e} = \eta_{e,0}(10^{8} \, {\rm  K}/T)^{2}$.
The rest of the constants, $C_{\cdots}$ and $D_{\cdots}$, reduce to integrals over the angular variables that can be done numerically.
Furthermore, when mutual friction is added to the superfluid model we find the following holds to about $4$ parts in $10^4$:
\beqa
\tau_v \rightarrow \tau_v + {A_{SF} \over {\cal S}^2 } \frac{T_8^2}{\Omega}
{\rm Re}\left(\sqrt{q_{SF} B + i \Omega /T_8^2}\right). \label{SFtau_vwithMF}
\eeqa
We find the constant $A_{SF}$ (which is positive and depends on ${\cal B}_n$) by fitting this formula to the results given by Eq.~(\ref{tauv}).

\section{Boundary Layer Damping Times and the Critical Angular Velocity}
\label{sectionIII}

One goal this study is to deterine how magnetic fields change the
cricital angular velocity needed for the onset of
the $r$-modes instability via their effect on the MVBL damping rate in accreting neutrons stars,
such as those in LMXBs.
Bildsten and Ushomirsky \cite{bu} predicted that magnetic fields would be important
when $B \ge 10^{11} \, {\rm G} \, (10^{8} \, {\rm K} /T)$ and this was confirmed by Mendell \cite{mendell2001}
for the ordinarly-fluid model.
However, Cumming, Zweibel, and Bildsten \cite{czb} have shown that buried fields in the crusts of LMXBs are less
than $10^{11} {\rm G}$, while typical LMXBs have external fields that are about $10^{8-9} {\rm G}$.
A minor coding error in Mendell \cite{mendell2001} caused the MVBL damping times to come out about 53\%
percent too small and the critical angular velocities to come out about 11\%
too large, without changing the qualitative results.
In this section we present corrected results for the ordinary-fluid model
and extend those results to the superfluid model.

As shown in previous studies, gravitational radiation emitted by the $r$-modes always
tends to drive these modes unstable.  The onset of the instability occurs when the dissipation rate equals the
gravitational-radiation growth rate, $\tau_{GR}$.  Thus, for our models we calculate the critical angular velocity for the onset
of the instability by solving
\beqa
\tau_v = \tau_{GR}, \label{tauvEQtaugr}
\eeqa
where $\tau_v$ is given by Eq.~(\ref{tauv}) [see also Eqs.~(\ref{OFtau_v})-(\ref{SFtau_vwithMF})]
and $\tau_{GR}$ for the $r$-modes is given by
\beqa
\tau_{GR} = \tilde{\tau}_{GR} \left ( {\Omega_o \over \Omega} \right )^{2m + 2} , \label{taugr}
\eeqa
where $\Omega_o  = \sqrt{\pi G \bar{\rho}}$, $G$ is the Newtonian gravitational constant,
and $\bar{\rho}$ is the average density of the star
(see references \cite{andersson,fried-morsink,lom,aks,owen-etal}).

Attention is restricted to the case used in previous studies: the $m = 2$ $r$-mode for a $1.4 M_\odot$ $n = 1$
polytrope, as described in, e.g., Lindblom, Mendell, and Owen \cite{lmo}, and Lindblom and Mendell \cite{lm2000}.
Adoption of this model allows easy comparison with previous studies. (Note also that the $m = 2$ case is the
most susceptible to the $r$-mode instability.)
For this case, $\kappa_0  = 2/3$, the stellar radius is $12.53 \, {\rm km}$, and $\Omega_o = 8413 \,{\rm s}^{-1}$.
Note that the
maximum angular velocity, where mass shedding occurs at the equator, is roughly $2\Omega_o/3$ (for any equation of state).
The density at the crust-core boundary is given approximately by
$1.5 \times 10^{14} \,{\rm g/cm}^3$ (see \cite{douchinhaensel,haensel}).
Using this density, the characteristic gravitational-radiation growth time
is $\tilde{\tau}_{GR} = 4.25 \,{\rm s}$ \cite{lou},
the core radius is $R_c = 11.01 \, {\rm km}$, the proton density
at the boundary is $\rho_p = 6.6 \times 10^{12} \,{\rm g/cm}^3$, and
the superfluid entrainment factor is $\gamma = 1.90498$ (for
an entrainment parameter of $0.04$, as defined in Lindblom and Mendell \cite{lm2000}).
In general, the dimensionless mutual friction
coefficient is given by \cite{mendell1991b,lm2000},
\beqa
{\cal B}_n = 5.1 \times 10^{-5}{ (\gamma - 1)^2 \rho_n \rho_p^{7/6}
\over
\rho^{1/2}(\gamma\rho_n + \rho_p)^{3/2}
} . \label{Bneqn}
\eeqa
Thus, at the crust-core boundary the mutual friction coefficient is ${\cal B}_n = 9.5 \times 10^{-5}$.
Furthermore, the energy per unit length of a proton vortex is given in Mendell \cite{mendell1991b} and in a more
convenient form in Mendell \cite{mendell98}. However, even the latter form depends on uncertain parameters that
involve the superconducting transition temperature and the entrainment factor.  Here, all these
uncertainties will be put into a single parameter $\varepsilon$.  In this case, $\varepsilon_p$
can be written as:
\beqa
\varepsilon_p = \varepsilon \rho_{p, 6.6e12}.
\eeqa
For typical neutron star numbers
\beqa
\varepsilon \cong 1.4 \times 10^6 \, {{\rm erg} \over {\rm cm}},
\eeqa
though its exact value is uncertain.
Values for the viscosities are also needed.  The electron and ordinary-fluid viscosities are given by \cite{cutler-lind}
\beqa
\eta_e = \biggl (1.35 \times 10^{19} {{\rm g} \over {\rm cm} \cdot {\rm s}}\biggr ) \rho_{1.5e14}^2 T_8^{-2},
\label{etae}
\eeqa
\beqa
\eta = \biggl (2.73 \times 10^{14} {{\rm g} \over {\rm cm} \cdot {\rm s}} \biggr ) \rho_{1.5e14}^{9/4} T_{10}^{-2}.
\label{etao}
\eeqa
Note that we define $\rho_{1.5e14} = \rho/(1.5 \times 10^{14} \, {\rm g} \cdot {\rm cm}^{-3})$,
$T_8  = T/(10^8 \, {\rm K})$ (and so on for other numeric subscripts thoughout the rest of this paper).

Finally, one important simplification must be pointed out.
As discussed at the end of the last section, it is convenient to choose
$B = B_r =$ constant. As already explained, this is unrealistic
but should give the same qualitative results as a realistic field (see Mendell \cite{mendell2001}).

Before finding the results for the critical angular velocity,
it is easy to show that magnetic effects will be
important for the superfluid model for the typical fields found in LMXBs.
If the magnetic terms are larger than the viscous terms
in Eq.~(\ref{kpmeqn}) then magnetic effects on the MVBL
length-scales will be important. Thus, the condition for this is
\beqa
{V_{\rm CV}^2 \over \kappa \Omega} \ge {\eta_e \over \rho_p}, \label{VCVlimiteqn}
\eeqa
Using Eq.~(\ref{SecIIVCVeqn}) in Eq.~(\ref{VCVlimiteqn})
yields the following lower bound on the magnetic field, such that magnetic effects dominate
the MVBL properties:
\beqa
B \ge (3.8 \times 10^9 \,{\rm G}) \kappa \Omega_{600\pi} \rho_{1.5e14}^2 T_8^{-2} \rho_{p, 6.6e12}^{-1} \varepsilon_{1.4e6}^{-1},
\label{Bsuperlimit}
\eeqa
Thus, in the superfluid model radial magnetic fields $\sim 10^9 \, {\rm G} \, (10^{8} \, {\rm K} /T)^2 $,
not untypical of the fields of LMXBs, can affect the MVBL. Though, as we will show,
the effects of mutual friction tend to counteract the magnetic effects.

For the physical parameters and restrictions just discussed, we
solve Eq.~(\ref{tauvEQtaugr}) for the critical angular velocity $\Omega_c$.
The temperature dependence of the critical angular velocity is presented in
Figs.~\ref{FigOrdBflds}-\ref{FigSupPinPBflds}.
In these figures, the horizontal dashed line corresponds to $2\Omega_o/3$,
which is the approximate maximum angular velocity for which mass-shedding occurs.  Thus, the regions above this
line are unphysical, and are kept only to illustrate the dependence of the curves for the range
of magnetic field magnitudes typically found in neutron stars.  Note that we extend the ordinary fluid
curves into the superfluid region ($T \le 10^9 {\rm K}$) and vice versa, given the uncertainties in the
superfluid transition temperature.

\bfig \centerline{\psfig{file=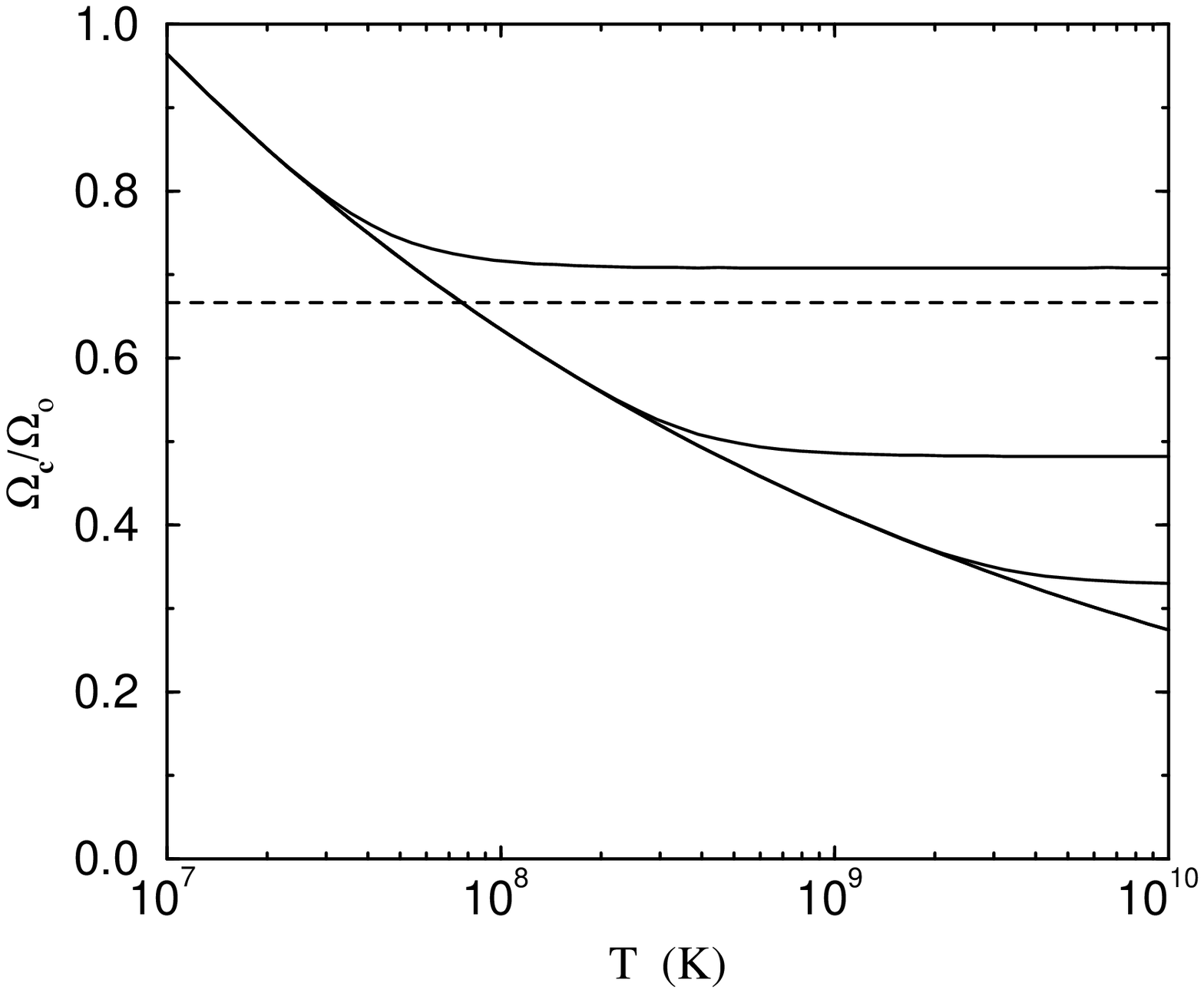,height=2.4in}} \vskip 0.3cm
\caption{Temperature dependence of the critical angular velocities for ordinary-fluid neutron stars
for $B = 0$, $10^{9}$ (indistinguishable from $B = 0$), $ 10^{10}$, $10^{11}$, and $10^{12}$ Gauss,
from bottom to top.\label{FigOrdBflds}} \efig

\bfig \centerline{\psfig{file=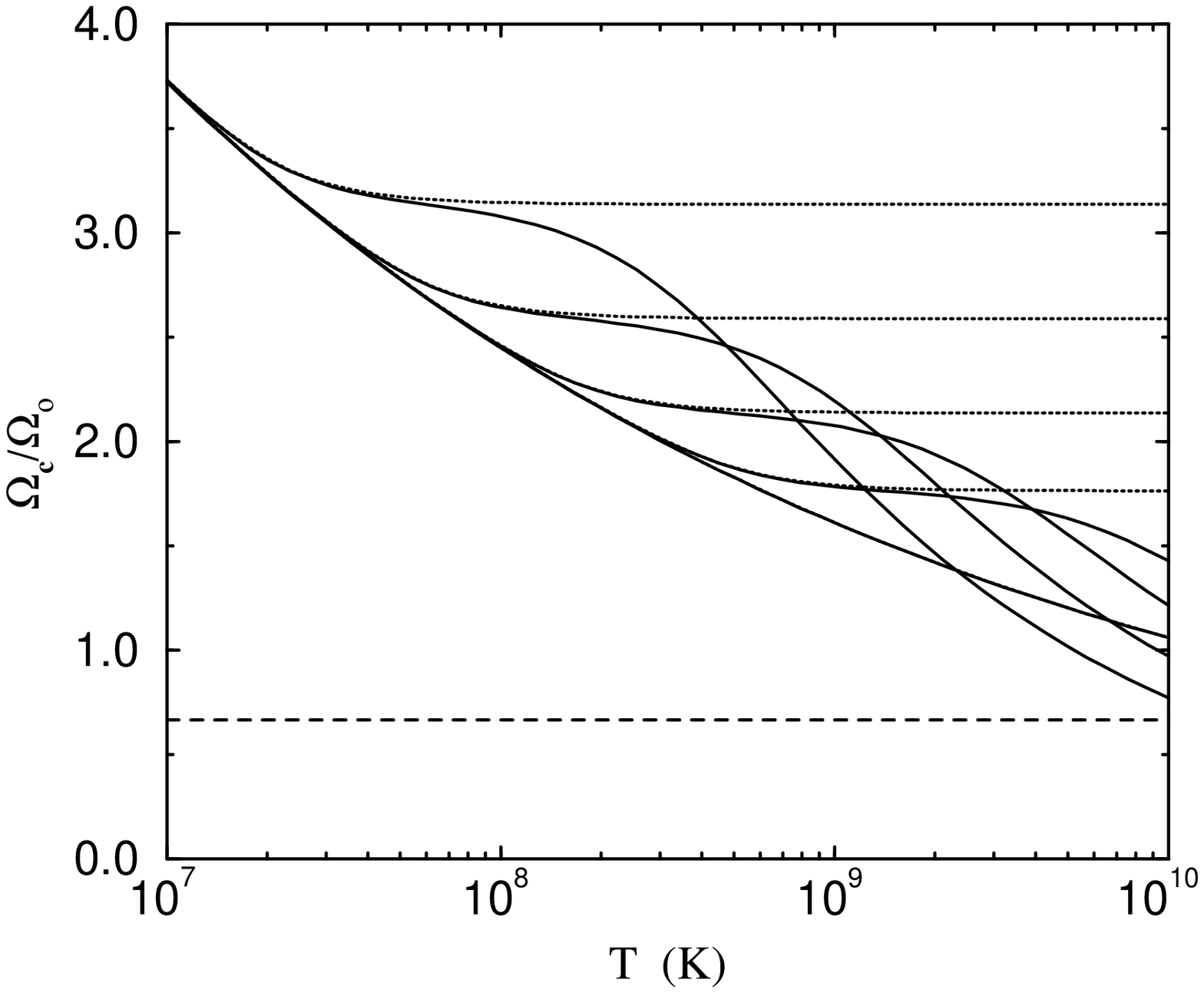,height=2.4in}} \vskip 0.3cm
\caption{Temperature dependence of the critical angular velocities for superfluid neutron stars
with pinned neutron vortices for $B = 0$, $10^{9}$, $ 10^{10}$, $10^{11}$,
and $10^{12}$ Gauss, for dotted lines from bottom to top. Note that the dotted lines correspond to no
mutual friction, while solid curves that diverge from these include mutual friction. \label{FigSupPinNBflds}} \efig

\bfig \centerline{\psfig{file=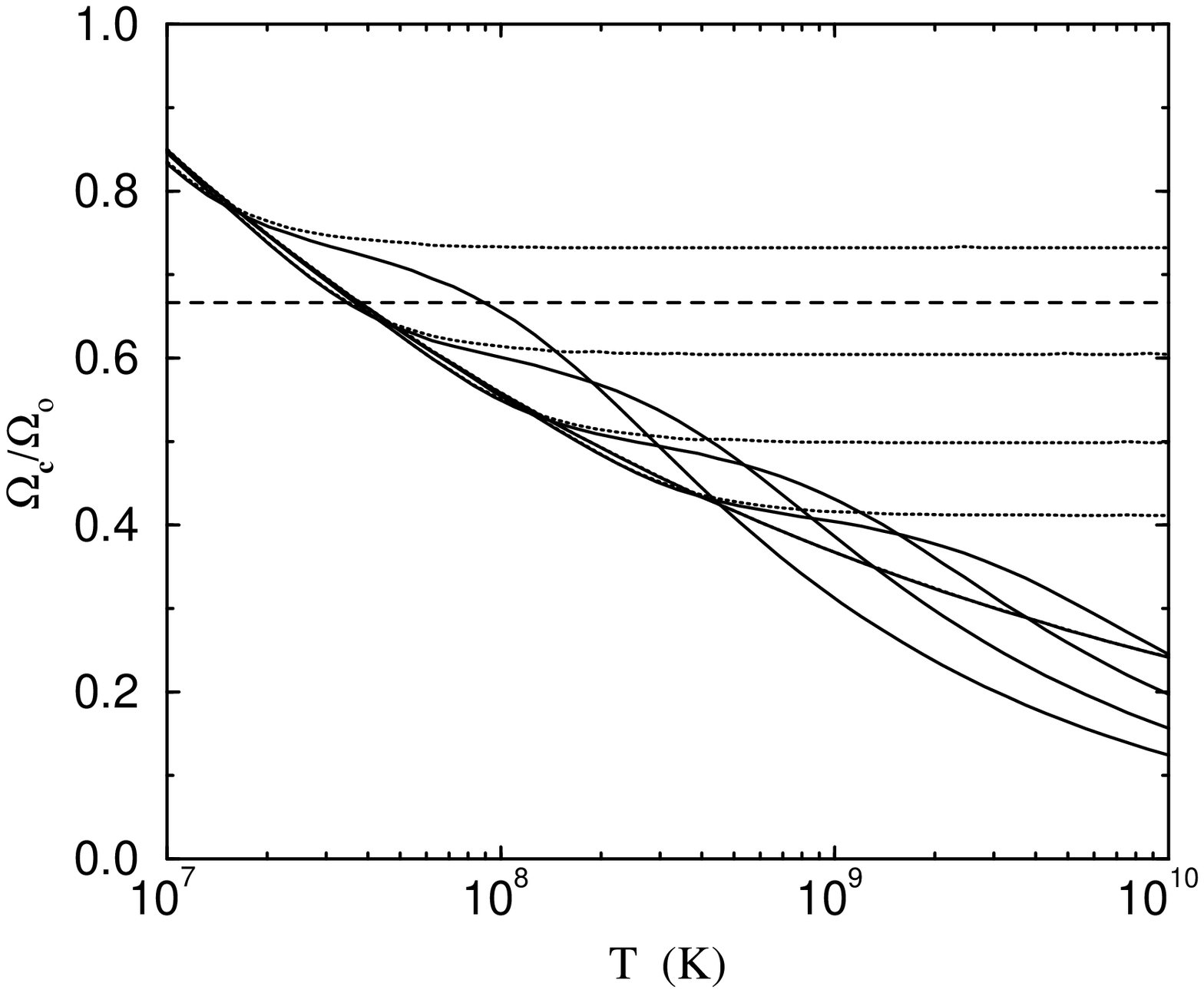,height=2.4in}} \vskip 0.3cm
\caption{Temperature dependence of the critical angular velocities for superfluid neutron stars
with pinned proton vortices for $B = 0$, $10^{9}$, $ 10^{10}$, $10^{11}$,
and $10^{12}$ Gauss, for dotted lines from bottom to top. Note that the dotted lines correspond to no
mutual friction, while solid curves that diverge from these include mutual friction. \label{FigSupPinPBflds}} \efig

\bfig \centerline{\psfig{file=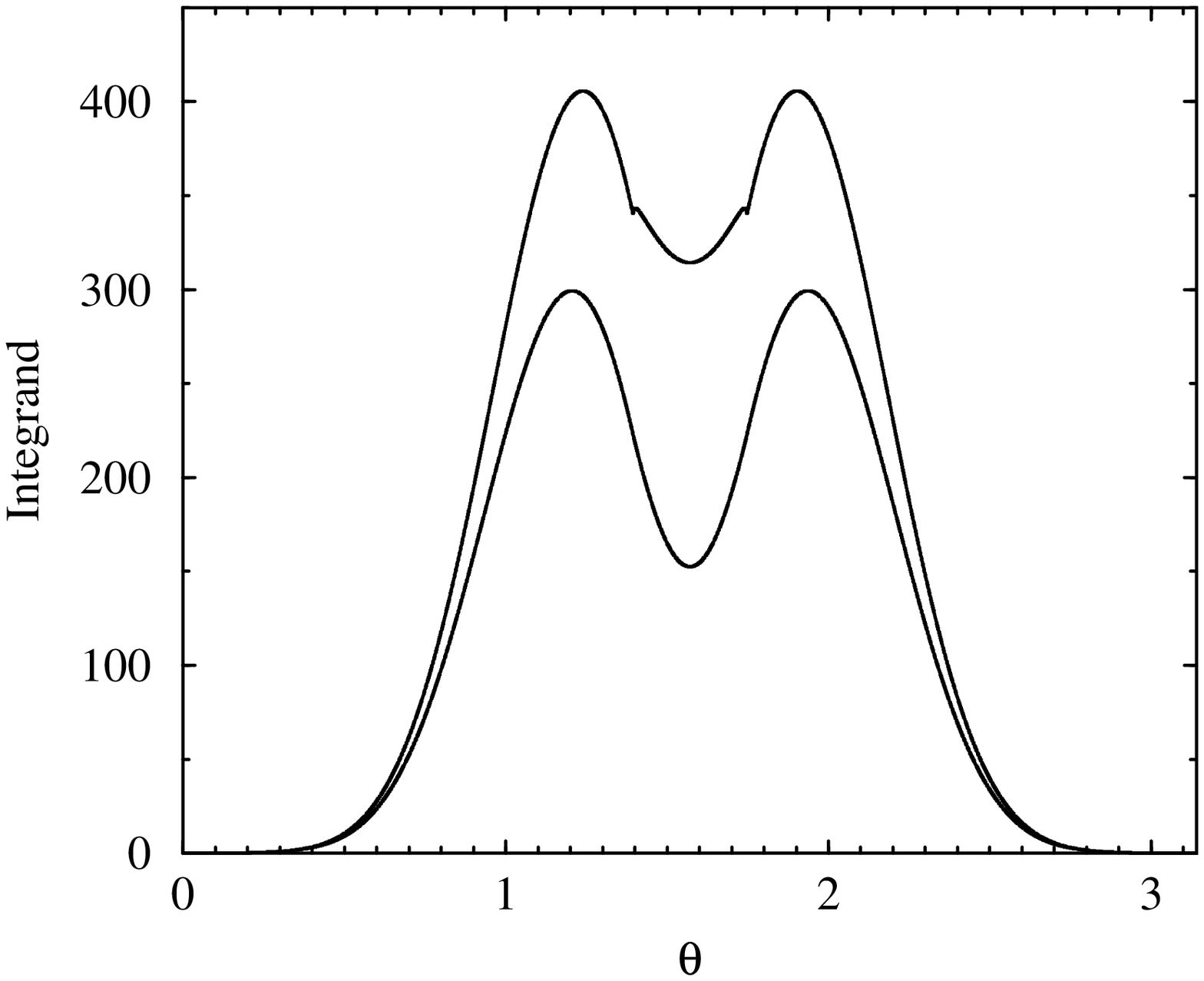,height=2.4in}} \vskip 0.3cm
\caption{Integrand that gives ${\cal I}$ for $T = 10^9 {\rm K}$, $ B = 10^{10} {\rm G}$,
including mutual friction (lower curve) and excluding mutual friction (upper curve). \label{FigIntegrands}} \efig

Figure~\ref{FigOrdBflds} shows the
temperature dependence of the critical angular velocities for ordinary-fluid neutron stars
for $B = 0$, $10^{9}$ (indistinguishable from $B = 0$), $ 10^{10}$, $10^{11}$, and $10^{12}$ Gauss,
from bottom to top. Overall, it is seen that magnetic fields increase the dissipation rate, raising
the critical angular velocity, and flatten out the curves.  The slip factor, ${\cal S}$, was set equal to $1$ for all these curves.
Smaller values of ${\cal S}$ lower these curves while preserving their shape.

Figures~\ref{FigSupPinNBflds} and \ref{FigSupPinPBflds} show results for the superfluid model for the
respective cases when just the neutron vortices are pinned or when just the proton vortices are pinned.
The dotted lines are for $B = 0$, $10^{9}$, $ 10^{10}$, $10^{11}$, and $10^{12}$ Gauss, from bottom to top
when mutual friction is ignored (${\cal B}_n = 0$), while the solid curves that diverge from these
include the effects of mutual friction. The slip factor, ${\cal S}$, was set equal to $1$ for all these curves.
Smaller values of ${\cal S}$ lower these curves while preserving their shape.
Thus, while many of these curves are far above the break-up velocity of the star, they can be moved
into the physically allowed region by choosing a small value for ${\cal S}$, which corresponds to
weak pinning of the vortices.  As in the ordinary-fluid model,
magnetic fields increase the dissipation rate, raising
the critical angular velocity, and flatten out the curves.  However, it is also seen
that mutual friction tends to counteract the magnetic effects for high temperatures
or high fields.

The results shown here were computing numerically using $\tau_v$ as given by Eq.~(\ref{tauv}).
However, we get the same results within a few tenths of a percent
using Eqs.~(\ref{OFtau_v})-(\ref{SFtau_vwithMF}).  For our models,
the constants in these equations are
$q_{OF} = 4.37 \times 10^{-16} \, {\rm s}^{-1} \cdot {\rm G}^{-2}$,
$C_{OF} = 1.41 \times 10^{-6} \, {\rm s}^{-1/2}$,
$D_{OF} = 1.56 \times 10^{-6} \, {\rm s}^{-1/2}$,
$q_{SF} = 7.41 \times 10^{-7} \, {\rm s}^{-1} \cdot {\rm G}^{-1}$,
while
$C_{SF} = 0.261 \, {\rm s}^{-1/2}$,
$D_{SF} = 0.243 \, {\rm s}^{-1/2}$,
$A_{SF} = 0.0150 \, {\rm s}^{1/2}$,
for pinned neutron vortices, and
$C_{SF} = 4.31 \times 10^{-5} \, {\rm s}^{-1/2}$,
$D_{SF} = 1.05 \times 10^{-4} \, {\rm s}^{-1/2}$,
for pinned proton vortices (we did not fit our data to determine $A_{SF}$ for this case).
Using these numbers and Eqs.~(\ref{OFtau_v})-(\ref{SFtau_vwithMF}), analytic expressions for $\tau_v$
for large $B$ are
\beqa
\tau_v = 33.9 \, {\rm s} \, {\cal S}^{-2} B_{12}^{-1},
\eeqa
for the ordinarly-fluid model, and
\beqa
&& \tau_v = 0.141 \, {\rm s} \, {\cal S}^{-2} B_{9}^{-1/2} \nonumber \\
&& \qquad + 4.85 \times 10^{-5} \, {\rm s} \, {\cal S}^{-2} T_8^2 B_{9}^{1/2} (\Omega_o/\Omega), \label{tauvanalyticSF}
\eeqa
for the supefluid model when neutron vortices are pinned.
Substituting these equations into Eq.~(\ref{tauvEQtaugr}) and using Eq.~(\ref{taugr}) gives
the following analytic expressions for the critical angular velocities
\beqa
{\Omega_c \over \Omega_o} = 0.71 {\cal S}^{1/3} B_{12}^{1/6},
\eeqa
for the ordinarly-fluid model,
while for the superfluid model with pinned neutron vortices we get
\beqa
{\Omega_c \over \Omega_o} = 1.76 {\cal S}^{1/3} B_{9}^{1/12} .
\eeqa
when the first term on the right in Eq.(\ref{tauvanalyticSF}) dominates, and
\beqa
{\Omega_c \over \Omega_o} = 9.74 {\cal S}^{2/5} T_8^{-2/5} B_{9}^{-1/10}. \label{OmegacritanalyticwithMF}
\eeqa
when the second term on the right in Eq.(\ref{tauvanalyticSF}) dominates.

We now explain why our results scale with $B$, $T$, and $\Omega$ as they do.
Note that in both the ordinary-fluid and superfluid models that the critical angular velocity
becomes temperature independent for high temperatures (excluding mutual friction).
However, one surprising result is that mutual friction actually lowers the damping rate.
To understand the various cases
we need to study the characteristic length-scales associated with the boundary layer.

The length-scales in the MBVL in the ordinary-fluid case have already been examined in Mendell \cite{mendell2001}.
Here we will consider the superfluid case.
The characteristic length-scales of the MVBL are determined by examining Eq.~(\ref{kpmeqn})
and the definitions of the length-scales given in Eqs.~(\ref{lamdefined}) and (\ref{ddefined}).
Dropping subscripts for the purposes of discussion, since the MVBL damping rate is controlled by terms of
the form $\eta |k|^2 d = \eta (4\pi/\lambda^2 + 1/d^2)d$, it would seem that the smallest length-scales
would determine the damping rate. However, as found by Mendell \cite{mendell2001}, the ratio $d/\lambda$ turns out to be more important
when magnetic effects are important.
Also, only real frequencies, $\kappa \Omega$, will be considered.  This is valid either when the system is driven
at a real frequency, or when the imaginary part of the frequency is small and can be ignored.
(This is true for our models except near certain special angles, as will be explained.)
Also, to further simplify the discussion, it will always be assumed that $\kappa > 0$.

First, consider Eq.~(\ref{kpmeqn}) when $B$ is less than
the lower bounds given in Eq.~(\ref{Bsuperlimit}). The viscous term dominates, and the
boundary layer thicknesses and wavelengths are given to lowest order by
\beqa
d_{\eta} = {\lambda_{\eta} \over 2\pi} = {1 \over |K_\pm|} \sqrt{{2 \eta_e \over \Omega \rho_p}} . \label{detaeqn}
\eeqa
For typical neutron star numbers these length-scales are
\beqa
d_{\eta} = {\lambda_{\eta} \over 2\pi} = {47 \,{\rm cm} \over |K_\pm|} \rho_{1.5e14} T_8^{-1}
\Omega_{600\pi}^{-1/2}\rho_{p, 6.6e12}^{-1/2}. \label{valdetaeeqn}
\eeqa
The effect of mutual friction is to only add small corrections to the above, and thus
does not matter in this case.
This equation agrees in form with the standard result,
but differs from previous results by numerical factors of
order unity due to the scaling with $|K_\pm|$, and more importantly, because
the proper scaling with the proton mass density is given here for the superfluid case,
for the first time.  This increases the VBL thickness, and consequently the damping time,
compared to previous estimates for the superfluid case,
e.g., by a factor of about $4.5$ compared to the
calculation given by Lindblom, Owen, and Ushomirsky \cite{lou}.

Mutual friction does matter for the case when $B$ is greater than
the lower bounds given in Eq.~(\ref{Bsuperlimit}).  However, we first consider the case
when mutual friction can be ignored.  In this case $K_\pm$ is either purely real
or purely imaginary depending on the angle $\theta$.
Taylor expanding Eqs.~(\ref{kpmeqn}) for this case the wave number is given by
\beqa
k_\pm = K_\pm \sqrt{{\kappa \Omega^2 \over V_{\rm CV}^2}}
\biggl [1 - {i \over 2} {\eta_e \over \rho_p} {\kappa \Omega \over V_{\rm CV}^2} \biggr ].
\label{Taylorkpm}
\eeqa
For the case of real $K_\pm$ the boundary layer length-scales to lowest order are
\beqa
{\lambda_{B} \over 2 \pi} = {1 \over |K_\pm| \kappa^{1/2}}
{V_{\rm CV} \over \Omega},
\label{lamBeqn}
\eeqa
which is basically the distance a cyclotron-vortex wave travels in one rotation, and
\beqa
d_{B/\eta} = {2 \over |K_\pm| \kappa^{3/2}}
{V_{\rm CV}^3 \rho_p \over \Omega^2 \eta_e}.
\label{dBetaeqn}
\eeqa
The subscripts indicate whether these quanties depend on purely magnetic, or a ratio of magnetic to viscous
quantities.  Substituting in values for the parameters gives
\beqa
{\lambda_{B} \over 2 \pi} = {33 \,{\rm cm} \over |K_\pm| \kappa^{1/2}}B_{3.8e9}^{1/2}\Omega_{600\pi}^{-1}
\varepsilon_{1.4e6}^{1/2},
\label{vallamBeqn}
\eeqa
\beqa
d_{B/\eta} = {65 \,{\rm cm} \over |K_\pm| \kappa^{3/2}}B_{3.8e9}^{3/2}\Omega_{600\pi}^{-2} T_8^{2}
\varepsilon_{1.4e6}^{3/2}{\rho_{p, 6.6e12} \over \rho_{1.5e14}^{2}}.
\label{valdBetaeqn}
\eeqa
For the case of imaginary $K_\pm$ the roles of $\lambda$ and $d$ become interchanged. Thus, for this case,
the results are as in Eqs.~(\ref{lamBeqn})-(\ref{valdBetaeqn}) with
\beqa
&& d \rightarrow {\lambda \over 2 \pi} , \nonumber \\
&& \lambda \rightarrow 2 \pi d.
\eeqa
It is apparent that for large $B$ that the MVBL damping rate is dominated by terms of the type
\beqa
{ 1 \over \tau_v} \sim \eta |k|^2 d \sim \eta {d_{B/\eta} \over \lambda_{B}^2} \sim B^{1/2}.
\eeqa
This explains the regions where the damping rate becomes independent of the temperature
and angular velocity, and the scaling of $\tau_v$ with $B^{-1/2}$ in the first
term on the right side Eq.~(\ref{tauvanalyticSF}).
Physically, the reason that magnetic fields increase the damping rate is
that while magnetic forces increase the size of the boundary layer many wavelengths
of cyclotron-vortex waves fit into the boundary layer thickness ($\lambda < d$) and
there is a large amount of shear associated with these waves.  (Similar explanations
are given in Mendell \cite{mendell2001} for the ordinary-fluid case.)

Attention will now be given to cases that occur near certain special angles in reference to Eq.~(\ref{Keqn}).
The angles are defined by
\beqa
\kappa \pm 2 {\rm cos}X_\pm = 0, \label{thetaA}
\eeqa
\beqa
\kappa \pm 2 \gamma {\rm cos}Y_\pm = 0, \label{thetaB}
\eeqa
\beqa
\left ( {\rho_n \gamma + \rho_p \over \rho} \right ) \kappa \pm 2 \gamma {\rm cos}Z_\pm = 0. \label{thetaC}
\eeqa
The latter two of these angles are important to understanding the effects of mutual friction
(and mutual friction smooths out singular behavior near these angles).
For the models used in this paper, these angles are located at: $X_- = 1.23$, $X_+ = 1.91$, $Y_- = 1.39$,
$Y_+ = 1.74$, $Z_- = 1.24$, and $Z_+ = 1.90$.

First, for $\theta$ near $X_\pm$, the small imaginary part of $\kappa$ becomes important
(in both the ordinary-fluid and superfluid models).  As explained in Mendell \cite{mendell2001},
to understand this, make the replacement $\kappa \rightarrow \kappa + i/(\Omega \tau)$, and note that
$\Omega \tau \gg 1$ for the situations of interest.
It is seen that $k_\pm \propto \sqrt{i/(\Omega \tau)}$ and that the
angular derivative $\partial k_\pm / \partial \theta$ becomes
large near $X_\pm$ (infinite at $X_\pm$ when the imaginary part of $\kappa$ is ignored).
However, as shown in Mendell \cite{mendell2001}, these regions in the ordinary-fluid model
make little contribution to the VBL damping rate.
In the superfluid model, the proximity of $X_\pm$ to $Z_\pm$ tends to narrow
the singular behavior near $X_\pm$.  Thus, correcting for the small
imaginary part of $\kappa$ near $\theta = X_\pm$ is not done in this paper,
nor has it been made in previous studies.

Now consider the case when mutual friction is included. Its effects are easiest to understand
near the angles $Y_\pm$ and $Z_\pm$, which are special only in the superfluid model.
The imaginary part of $\kappa$ could also be important near these angles,
but note that mutual friction effects are more important
as long as $\tau > 1/(2\gamma\Omega{\cal B}_n)$.
Thus, mutual friction effects dominates near $Y_\pm$ and $Z_\pm$
when the damping time is greater than the lower bound
$\tau > 2.7 \, {\rm s} / (\gamma \Omega_{600\pi} {\cal B}_{n,1.0e-4})$.
It is found that this lower bound is valid for the results presented in this paper.
Also, recall that mutual friction is only important for the case $B$ is greater than
the lower bound given in Eq.~(\ref{Bsuperlimit}).
With these caveats in mind, first consider the effects of mutual friction when $\theta$ is near $Y_\pm$.
One can show that  the characteristic
MVBL length-scales are
\beqa
d_{B/{\cal B}_n} = {\lambda_{B/{\cal B}_n} \over 2\pi} =  \sqrt{{2 \over \kappa}}
\sqrt{{\rho_n \rho_p \over 2 \rho^2 {\cal B}_n}}
{V_{\rm CV} \over \Omega},
\eeqa
which for typical neutron star numbers is
\beqa
&& d_{B/{\cal B}_n} = {\lambda_{B/{\cal B}_n} \over 2\pi} =
{670 \, {\rm cm} \over \sqrt{\kappa}} B_{3.8e9}^{1/2} \Omega_{600\pi}^{-1} \varepsilon_{1.4e6}^{1/2}
\nonumber \\
&& \qquad \qquad \times {\rho_{n,1.434e14}^{1/2} \rho_{p,6.6e12}^{1/2} \over \rho_{1.5e14}}
{\cal B}_{n, 1.0e-4}^{-1/2} . \label{valdBBneqn}
\eeqa
Next, for $\theta = Z_\pm$, the
MVBL length-scales to lowest order are
\beqa
d_{B \cdot {\cal B}_n} =
{\lambda_{B \cdot {\cal B}_n} \over 2\pi} =
{\gamma \over (\gamma - 1)} \sqrt{{2 \over \kappa}}
\sqrt{{2 \rho^2 {\cal B}_n \over \rho_n \rho_p}}
{V_{\rm CV} \over \Omega},
\eeqa
which for typical neutron star numbers is
\beqa
&& d_{B \cdot {\cal B}_n} = {\lambda_{B \cdot {\cal B}_n} \over 2\pi} =
{(3.2 \, {\rm cm} ) \gamma \over (\gamma - 1) \kappa^{3/2}}
{B_{3.8e9}^{1/2} \over \Omega_{600\pi}}
\varepsilon_{1.4e6}^{1/2}
\nonumber \\
&& \qquad \qquad \times {\rho_{1.5e14} \over \rho_{n,1.434e14}^{1/2} \rho_{p,6.6e12}^{1/2}}
{\cal B}_{n, 1.0e-4}^{1/2} . \label{valdBdotBneqn}
\eeqa
We see that the length-scales again depend on the distance a cyclotron-vortex wave travels in one rotation, but
that $d = \lambda/ 2\pi$.  Thus we do not have many wavelengths within the boundary layer, but only one wavelenth.
We see that the length-scales near $Z_\pm$ are
similar to the case when mutual friction is ignored.
However, near $Y_\pm$ the length-scales are much larger
than in the case when mutual friction is ignored. This
reduces the dissipation rate and is the main reason that mutual friction reduces the critical
angular velocities as shown in Figures~\ref{FigSupPinNBflds} and \ref{FigSupPinPBflds}.
Also note, that when
$d = \lambda/2\pi$ the dissipation varies as $\eta |k|^2 d \sim \eta/d$, and so
the damping time varies as $d / \eta $.  For $d$ in Eqs.~(\ref{valdBBneqn}) and
(\ref{valdBdotBneqn}) we see that this implies $\tau_v \sim B^{1/2} T^2 / \Omega$.
Thus, we have explained the scaling of the mutual friction correction term
with $B$, $T$, and $\Omega$ found empirically
in Eq.~(\ref{SFtau_vwithMF}) and used
in Eqs.(\ref{tauvanalyticSF}) and (\ref{OmegacritanalyticwithMF}).
The overall effect of mutual friction is shown in Fig.~\ref{FigIntegrands}, which
again shows that mutual friction lowers the dissipation rate, and smooths out
singular behavior near the special angles considered here.

\section{Cyclic Evolution of Accreting Neutron Stars}
\label{sectionV}

We now turn our attention to the differential equations governing the spin cycles of accreting neutron stars.
In the context of the $r$-modes, these cycles have been discussed
by Levin \cite{levin}, Anderson, {\it et al.} \cite{ajks}, Wagoner, Hennawi, and Liu \cite{whl}, and Heyl \cite{heyl}.
Previously Wagoner \cite{wagoner84} suggested that the spins of these stars could be limited by gravitational-radiation
from an unstable mode, while Bildsten \cite{bildsten} proposed an alternative model in which the gravitational-radiation
is generated due to an accretion-induced asymmetry of the star.

First, as in Owen {\it et al.} \cite{owen-etal}, we define the dimensionless $r$-mode amplitude $\alpha$ as the
maximum perturbed fluid velocity at the equator of the star divided by $\Omega R$.
(Note that we do make corrections in this section for the fact the $r$-mode is confined to just the core
of the star but that this does not change the definition of $\alpha$.)
Second, following the analysis of
Wagoner, Hennawi, and Liu \cite{whl} (which builds on the work of Owen {\it et al.} \cite{owen-etal} and
Levin \cite{levin}),
the following equations govern the evolution of the $r$-mode amplitude, $\alpha$, the angular velcocity, $\Omega$,
and the temperature $T$:
\beqa
{1 \over \alpha} {d\alpha \over dt} = F_g - F_v + F_g K_c \alpha^2 - {1 \over 2} F_a, \label{alphaevolve}
\eeqa
\beqa
{1 \over \Omega} {d\Omega \over dt} = -2 F_g K_c \alpha^2 + F_a , \label{Omegaevolve}
\eeqa
\beqa
C(T) {d T \over dt} = W_{\rm diss} + K_n \dot{M} c^2 - L_\nu(T). \label{Tempevolve}
\eeqa
These equations, of course, are highly simplified in that they ignore differential rotation, assume infinite heat
conduction, and so forth, but give a qualitative overview of how the evolution should proceed.

We now discuss the various terms in these equations.  Heating due to dissipation is given by
\beqa
W_{\rm diss} = \tilde{J} M R^2 \Omega^2 \alpha^2  F_v . \label{Wdiss}
\eeqa
The constant $\tilde{J}$  is defined in Eq.~(3.4)
of Owen {\it et al.} \cite{owen-etal} and relates to the canonical angular momentum of the $r$-modes.
Here we recompute this constant for $r$-modes confined to just the core of the star and find its value is $\tilde{J} = 0.01255$.
The constant $K_c$, called $Q$ in Owen {\it et al.} \cite{owen-etal} and defined below their Eq.~(3.4),
is related to $\tilde{J}$ and the moment of inertia the star (which is unchanged even when the $r$-modes are confined to the core).
However, since $\tilde{J}$ changes, $K_c$ also changes, and we find $K_c = 0.072$.
Next, note that Wagoner, Hennawi, and Liu \cite{whl} define $1 - K_j$ as the fraction of the canonical $r$-mode
angular momentum that
contributes to the total physical angular momentum of the star. In these equations we have adopted the case that
the canonical angular momentum of the $r$-modes contributes zero physical angular momentum to the star, i.e., $K_j = 1$.
We find in our case, as Wagoner, Hennawi, and Liu \cite{whl} find for the cases they study, that the exact value of $K_j$ is not
important to the results since the $r$-modes saturate quickly and no more physical angular momentum can go into them,
if any at all does to begin with.  Furthermore, Levin and Ushomirsky \cite{lushell} show that the canonical angular momentum
of the $r$-modes contributes zero physical angular momentum to the star for a toy model, further justifying our
use of $K_j = 1$.
Next, note that the above equations are valid only for small $\alpha$.  This is OK even when the mode saturates as
long as the saturation amplitude is small. Several studies suggest the saturation amplitude is indeed small \cite{wumatarras,arrasetal}.
However, for comparison with earlier work we do consider a few cases where the saturation amplitude is $\alpha = 1$, and
for these few case only we instead use Eq.~(5) in Levin \cite{levin} to evolve $\Omega$ when $\alpha = 1$.
As in Wagoner, Hennawi, and Liu \cite{whl}, but using the MBVL damping rate which dominates for our models, we set
\beqa
F_g = {1 \over \tau_{GR} }, \label{Fgeqn}
\eeqa
where $\tau_{GR}$ is given by Eq.~(\ref{taugr}), while we set
\beqa
F_v = {1 \over \tau_v} , \label{Fveqn}
\eeqa
where $\tau_v$ is the MVBL damping time given by Eq.~(\ref{tauv}).
The inverse accretion time-scale, $F_a$, is related to the accretion rate, $\dot{M}$, by
\beqa
F_a = \left ( {1 \over 5 \times 10^{6} {\rm yr}} \right )
\left ( {\dot{M} \over 10^{-8} M_\odot / {\rm yr} } \right ), \label{Acceqn}
\eeqa
where the observed accretion rates in LMXBs are in
the range $10^{-11} M_\odot / {\rm yr} \le \dot{M} \le 10^{-8} M_\odot / {\rm yr}$ \cite{czb}.
We take the accretion rate to be constant.  Even if the accretion rate were not constant, while the $r$-modes are saturated the accretion rate
has little effect on the evolution. Note that heating due to accretion is proportional to the constant $K_n = 1 \times 10^{-3}$ \cite{whl}.
Finally, in Eq.~(\ref{Tempevolve}) the heat capacity and neutrino cooling rates are given by \cite{levin,whl,wagonerpc},
\beqa
&& C = 1.47 \times 10^{38} {\rm erg} \cdot {\rm K}^{-1} T_8, \\
&& L_\nu = L_{\rm URCA}' T_8^8 + L_{\rm brem}' T_8^6,
\eeqa
for the ordinary-fluid case, and
\beqa
&& C = 7 \times 10^{36} {\rm erg} \cdot {\rm K}^{-1} T_8,  \\
&& L_\nu = L_{\rm brem}' T_8^6,
\eeqa
for the superfluid case. The constants $L_{URCA}'$ and $L_{brem}'$
are given by \cite{wagonerpc},
\beqa
&& L_{\rm URCA}' = 1.5 \times 10^{32} {\rm erg s}^{-1} , \\
&& L_{\rm brem}' = 1.4 \times 10^{30} {\rm erg s}^{-1}.
\eeqa

\bfig \centerline{\psfig{file=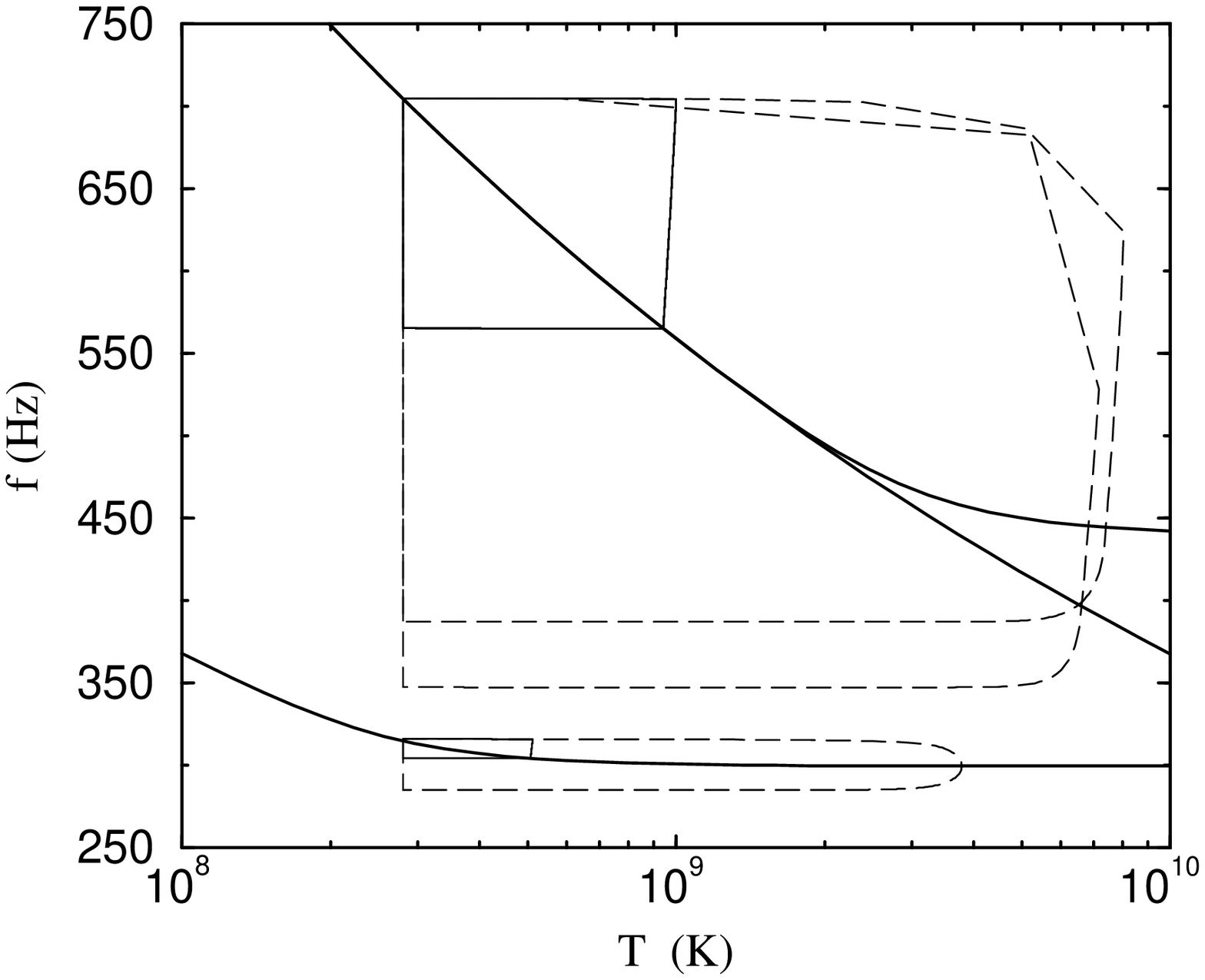,height=2.4in}} \vskip 0.3cm
\caption{Spin frequency vs. temperature evolution for ordinary-fluid neutron stars
for $B = 0 \, {\rm G}$ and $B = 10^{10} \, {\rm G}$ (both for ${\cal S} = 1.0$), and
for $B = 10^{11} \, {\rm G}$ (for ${\cal S} = 0.1$).
The curves are explained in the text. \label{FigEvolveOrdBFlds}} \efig

\bfig \centerline{\psfig{file=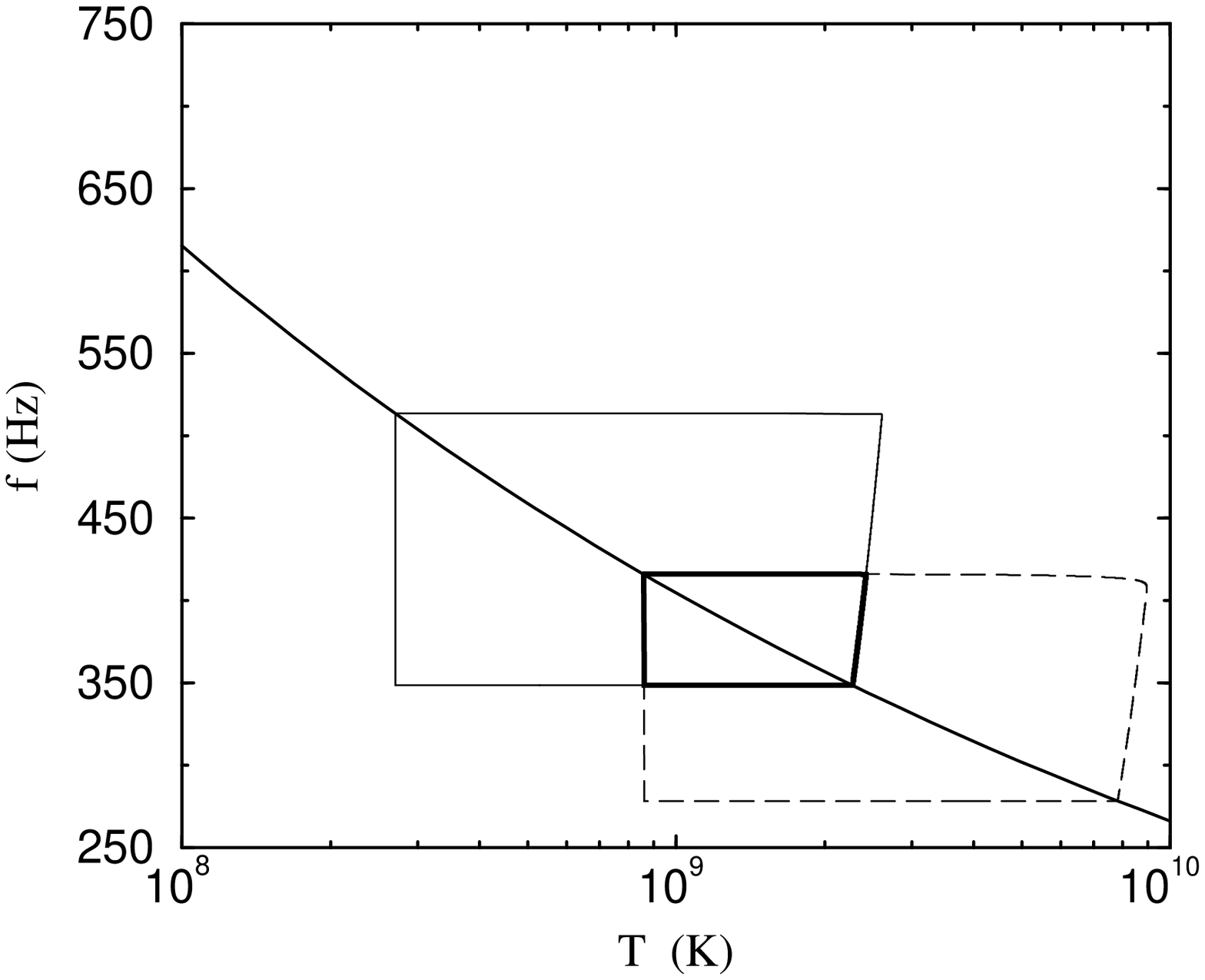,height=2.4in}} \vskip 0.3cm
\caption{Spin frequency vs. temperature evolution for superfluid neutron stars with pinned neutron vortices
and for $B = 0 \, {\rm G}$ and ${\cal S} = 0.01$. The curves are explained in the text. \label{FigEvolveSupB00}} \efig

\bfig \centerline{\psfig{file=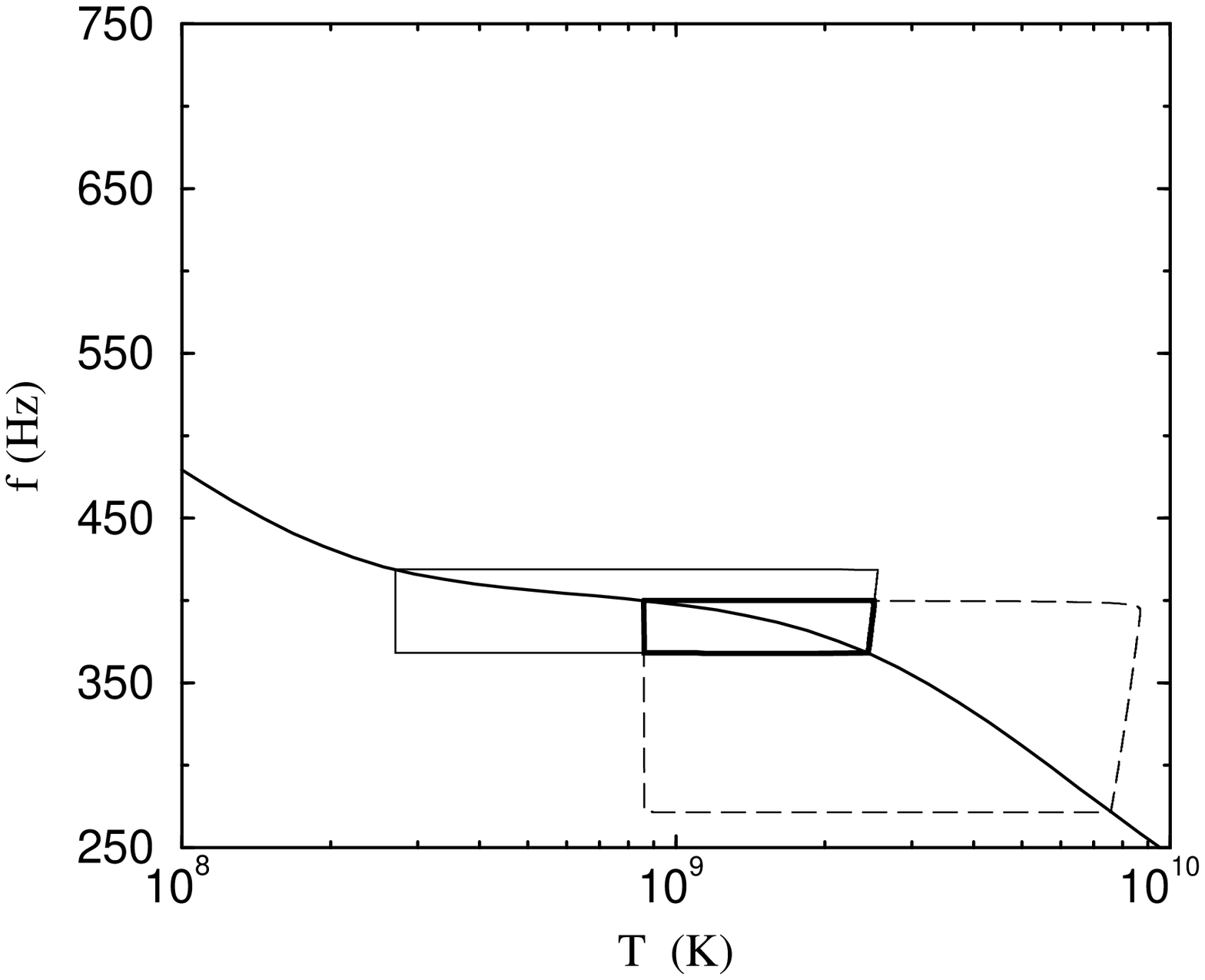,height=2.4in}} \vskip 0.3cm
\caption{Spin frequency vs. temperature evolution for superfluid neutron stars with pinned neutron vortices
and for $B = 10^{9}  \, {\rm G}$ and ${\cal S} = 0.005$. The curves are explained in the text. \label{FigEvolveSupB09}} \efig

\bfig \centerline{\psfig{file=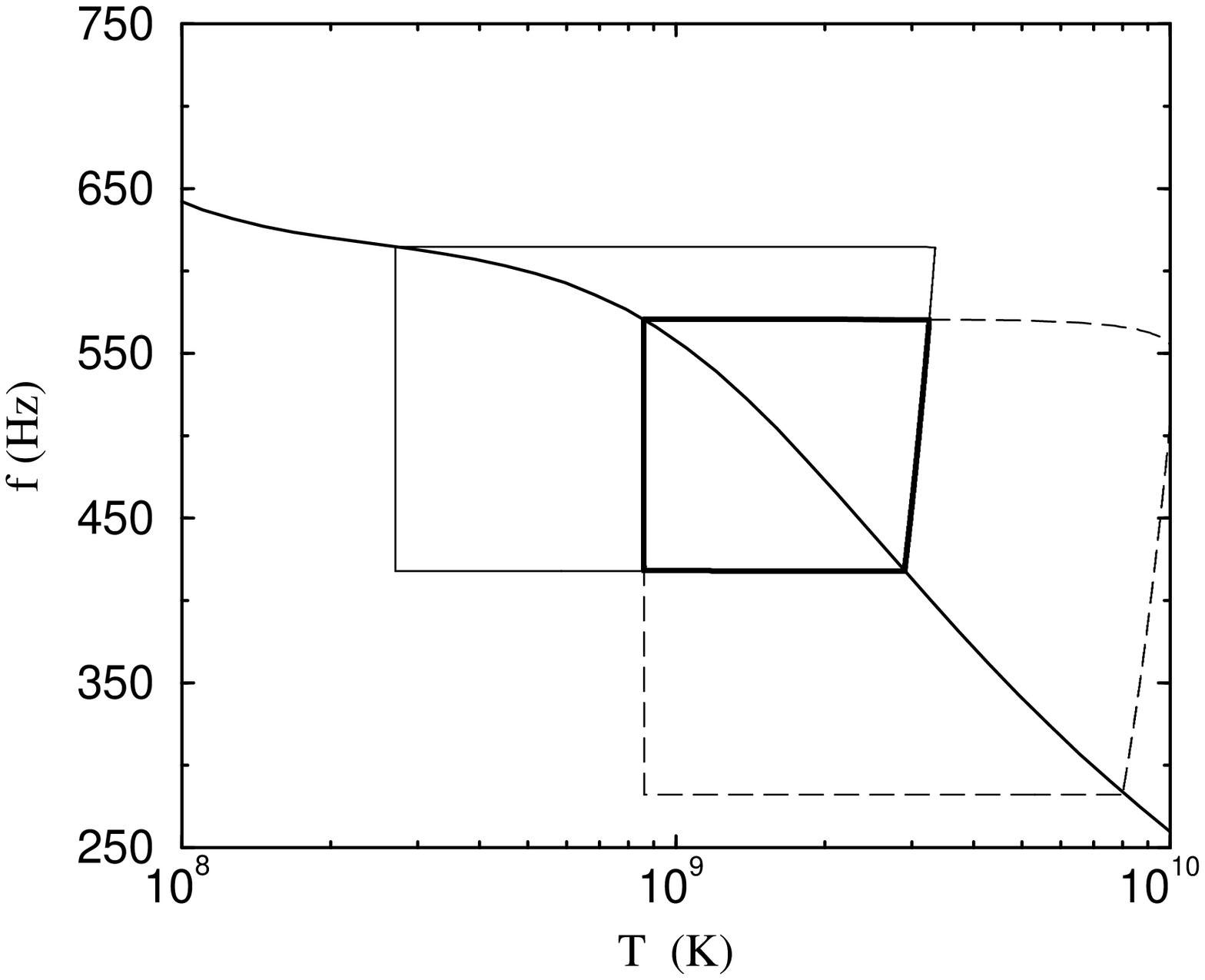,height=2.4in}} \vskip 0.3cm
\caption{Spin frequency vs. temperature evolution for superfluid neutron stars with pinned neutron vortices
and for $B = 10^{10} \, {\rm G}$ and ${\cal S} = 0.01$. The curves are explained in the text. \label{FigEvolveSupB10}} \efig

The results of evolving Eqs.~(\ref{alphaevolve})-(\ref{Tempevolve}) are the cyclic curves shown in Figs.~\ref{FigEvolveOrdBFlds}-\ref{FigEvolveSupB10}.
The monitonically decreasing smooth curve through the middle of each cycle is the critical angular velocity curve.
As time increases, a star progresses clockwise around the cycle.
Note that for each cycle there is a maximum and minimum angular velocity and a maximum and minimum temperature.
At the maximum angular velocity the $r$-mode becomes unstable at the minimum
temperature, i.e., we define $\Omega_{\rm max} = \Omega_c (T_{\rm min})$.
At the minimum angular velocity the $r$-mode becomes stable at the maximum
temperature, i.e., we define $\Omega_{\rm min} = \Omega_c (T_{\rm max})$.
We define the change in angular velocity by
$\Omega_{\rm max} - \Omega_{\rm min} = \Delta \Omega$.
We will describe some cycles as ``thin'' if $\Delta \Omega \ll \Omega_{\rm min}$,
otherwise we will describe the cycle as ``fat'' if $\Omega_{\rm min} \ll \Omega_{\rm max}$.
For each cycle the star goes through roughly 4 distinct stages.  Each stage corresponds
to one side of the roughly quadrulateral curve that corresponds to one cycle \cite{cycleshape}.
These stages have been previously discussed by Levin \cite{levin} and Anderson, {\it et al.} \cite{ajks} (see also Heyl \cite{heyl}).
Here we extend their results to include the superfluid case, magnetic effects, and thin cycles.  First we give typical numbers for our models.
After this we discuss the figures in detail.

The first stage corresponds to the vertical left side of the cycle, which corresponds to the star
spinning up by accretion with heating due to accretion balanced by
cooling due to neutrino emission $K_n \dot{M} c^2 = L_\nu(T)$.  Thus, the minimum temperture is related to the accretion rate by
\beqa
T_{\rm min} = 2.8 \times 10^{8} \, {\rm K} \, \left ( {\dot{M} \over 10^{-8} M_\odot / {\rm yr}} \right )^{1/8}, \label{TminOF}
\eeqa
for the ordinary-fluid case, and
\beqa
T_{\rm min} = 8.6 \times 10^{8} \, {\rm K} \, \left ( {\dot{M} \over 10^{-8} M_\odot / {\rm yr}} \right )^{1/6}, \label{TminSF}
\eeqa
for the superfluid case.
The time-scale for accretion to spin up the star is given by solving Eq.(\ref{Omegaevolve}) when $F_a$ dominates:
\beqa
&& t_{\rm spinup} = {1 \over F_a} \ln {\Omega_{\rm max} \over \Omega_{\rm min} } \nonumber \\
&& \qquad = 5 \times 10^{6} \, {\rm yr} \, \left ( {10^{-8} M_\odot / {\rm yr} \over \dot{M} } \right )
\ln \left ( {\Omega_{\rm max} \over \Omega_{\rm min}} \right ). \label{tspinup}
\eeqa
It will be useful to note that for thin cycles the spinup time is approximately,
\beqa
t_{\rm spinup} \cong
5 \times 10^{6} \, {\rm yr} \, \left ( {10^{-8} M_\odot / {\rm yr} \over \dot{M} } \right )
\left ( {\Delta \Omega \over \Omega_{\rm min}} \right ). \label{tspinupthin}
\eeqa

The second stage corresponds to the the top of the cycle. The beginning of this stage occurs when the $r$-mode becomes unstable and the
mode amplitude grows to a saturation amplitude $\alpha_{\rm sat}$.  This happens on the gravitational growth time-scale of
\beqa
&& t_{\rm grow} = {1 \over F_g} \ln {\alpha_{\rm sat} \over \alpha_{\rm min} } \nonumber \\
&& \qquad = 9.3 \, {\rm hr} \, \left ( {600 \pi / {\rm s} \over \Omega_{\rm max}} \right )^{6}
\ln \left ( { \alpha_{\rm sat} \over \alpha_{\rm min} } \right ). \label{tgrow}
\eeqa
As is done in Wagoner, Hennawi, and Liu \cite{whl}, we assume a residual $r$-mode amplitude is continuously excited by stochastic processes
and do not allow $\alpha$ to evolve below $\alpha_{\rm min} = 10^{-12}$.
As the mode grows, viscous dissipation heats the
star to a maximum temperture for which $\tilde{J} M R^2 \Omega^2 \alpha_{\rm sat}^2  F_v = L_\nu(T)$.  Thus, the maximum temperature is determined
by the saturation amplitude.  If we overestimate $F_v$ by replacing it with $F_g(\Omega_{\rm max})$ we find
\beqa
T_{\rm max} \lesssim 5.0 \times 10^{8} \, {\rm K} \, \left ( {\alpha_{\rm sat} \over 10^{-4}} \right )^{1/4}
\left ( { \Omega_{\rm max} \over 600 \pi / {\rm s}} \right ) , \label{TmaxOF}
\eeqa
for the ordinary-fluid case, and
\beqa
T_{\rm max} \lesssim 1.9 \times 10^{9} \, {\rm K} \, \left ( {\alpha_{\rm sat} \over 10^{-4}} \right )^{1/3}
\left ( { \Omega_{\rm max} \over 600 \pi / {\rm s}} \right )^{4/3} , \label{TmaxSF}
\eeqa
for the superfluid case.
Next, we can solve Eq.~(\ref{Tempevolve}) when the term with $F_v$ dominates to get the characteristic time for dissipation to heat the star
\beqa
t_{\rm heat} = {1 \over 2 W_{\rm diss}}
\left [ C(T_{\rm max}) T_{\rm max}  - C(T_{\rm min}) T_{\rm min}) \right ], \label{theat}
\eeqa
If we make the approximations $T_{\rm max} \gg T_{\rm min}$ and $F_v \lesssim F_g(\Omega_{\rm max})$, estimates
for this time-scale are
\beqa
t_{\rm heat} \sim 96 \, {\rm yr} \, \left ( {\alpha_{\rm sat} \over 10^{-4}} \right )^{-3/2}
\left ( { \Omega_{\rm max} \over 600 \pi / {\rm s}} \right )^{-6} , \label{theatOF}
\eeqa
for the ordinary-fluid case, and
\beqa
t_{\rm heat} \sim 69 \, {\rm yr} \, \left ( {\alpha_{\rm sat} \over 10^{-4}} \right )^{-4/3}
\left ( { \Omega_{\rm max} \over 600 \pi / {\rm s}} \right )^{-16/3} , \label{theatSF}
\eeqa
for the superfluid case.  Note that this time-scale dominates the top of the evolution cycle.

The third stage occurs when angular-momentum radiated away as gravitational radiation spins down the star.
Solving Eq.(\ref{Omegaevolve}) when the term with $F_g$ dominates gives
\beqa
t_{\rm spindown} = {\tilde{\tau}_{GR} \over 12 K_c \alpha_{\rm sat}^{2} }
\left [ \left ( {\Omega_o \over \Omega_{\rm min}} \right )^6 - \left ( {\Omega_o \over \Omega_{\rm max}} \right )^6 \right ] . \label{tspindown}
\eeqa
For thin cycles,
\beqa
&& t_{\rm spindown} \cong \nonumber \\
&& \qquad 7.4 \times 10^{5} \, {\rm yr} \, { \Delta \Omega \over \Omega_{\rm min} }
\left ( {10^{-4} \over \alpha_{\rm sat}} \right )^{2}
\left ( { 600 \pi / {\rm s} \over \Omega_{\rm min} } \right )^{6} , \label{tspindownthin}
\eeqa
while for fat cycles,
\beqa
t_{\rm spindown} \cong 1.2 \times 10^{5} \, {\rm yr} \, \left ( {10^{-4} \over \alpha_{\rm sat}} \right )^{2}
\left ( { 600 \pi / {\rm s} \over \Omega_{\rm min} } \right )^{6}  . \label{tspindownfat}
\eeqa
This stage dominates the right side of the evolution cycles.
Note that stages 2 and 3 begin at the same time.  Because they have very different durations they can be studied separately.

The fourth stage begins when the star spins down to the critical angular velocity curve.
The amplitude of the mode then decays away on the dissipation timescale
\beqa
&& t_{\rm damp} = {1 \over F_v} \ln {\alpha_{\rm sat} \over \alpha_{\rm min} } \nonumber \\
&& \qquad \sim 9.3 \, {\rm hr} \, \left ( {600 \pi / {\rm s} \over \Omega_{\rm min}} \right )^{6}
\ln \left ( { \alpha_{\rm sat} \over \alpha_{\rm min} } \right ), \label{damp}
\eeqa
where we have used $F_v \sim F_g(\Omega_{\rm min})$ during the damping to obtain the numerical estimate for
this time-scale.
The last stage is dominated by the time it takes the star to cool back to $T_{\rm min}$.  Solving Eq.~(\ref{Tempevolve}) when
the $L_\nu$ dominates gives
\beqa
&& t_{\rm cool} \cong { (10^{8} \, {\rm K})^8 \over 6 L_{\rm URCA}'}
\left [ {C(T_{\rm min}) \over T_{\rm min}^7} - {C(T_{\rm max}) \over T_{\rm max}^7} \right ] \nonumber \\
&& \qquad \sim 4.9 \times 10^{5} \, {\rm yr} \, \left ( { 10^8 \, {\rm K} \over T_{\rm min} } \right )^6 , \label{tcoolOF}
\eeqa
for the ordinary-fluid case, and
\beqa
&& t_{\rm cool} \cong { (10^{8} \, {\rm K})^6 \over 4 L_{\rm brem}'}
\left [ {C(T_{\rm min}) \over T_{\rm min}^5} - {C(T_{\rm max}) \over T_{\rm max}^5} \right ] \nonumber \\
&& \qquad \sim 4.0 \times 10^{6} \, {\rm yr} \, \left ( { 10^8 \, {\rm K} \over T_{\rm min} }\right )^4 , \label{tcoolSF}
\eeqa
for the superfluid case.
This stage dominates the bottom of the evolution curves.
Note that stage 4 really coincides with stage 1 of the next cycle.
Because they have somewhat different durations they can be studied separately.

Thus, as Levin \cite{levin}, Anderson, {\it et al.} \cite{ajks}, and Heyl \cite{heyl} conclude, we see that
the time that the $r$-mode spends saturated (and thus actively radiating potentially detectable gravitational waves)
is dominated by the spindown time-scale, while the time that the $r$-mode spends with minimum amplitude (and thus radiating
insignificantly) is dominated by the spinup time.  Thus, defining $r$ as the fraction of LMXBs in the active phase
gives $r \cong t_{\rm spindown}/t_{\rm spinup}$, as in Levin \cite{levin}.
For fat cycles, using Eqs.~(\ref{tspinup}) and (\ref{tspindownfat}),
\beqa
r \lesssim .072 \left ( {10^{-4} \over \alpha_{\rm sat}} \right )^{2}
\left ( { 500 \pi / {\rm s} \over \Omega_{\rm min} } \right )^{6}
\left ( { \dot{M} \over 10^{-8} M_\odot / {\rm yr} } \right ) , \label{rfat}
\eeqa
where $\ln (\Omega_{\rm max} / \Omega_{\rm min})$ was treated as a
factor of order unity.  For thin cycles, using Eqs.~(\ref{tspinupthin}) and (\ref{tspindownthin}),
\beqa
r \lesssim  .44 \left ( {10^{-4} \over \alpha_{\rm sat}} \right )^{2}
\left ( { 500 \pi / {\rm s} \over \Omega_{\rm min} } \right )^{6}
\left ( { \dot{M} \over 10^{-8} M_\odot / {\rm yr} } \right ) . \label{rthin}
\eeqa
The first of these ratios is the same as what Levin \cite{levin} found, though scaled differently here.
Note that for these equations we have scaled $\Omega_{\rm min}$ with the observed minimum value
$250 \, {\rm Hz}$, rather than a typical value of $300 \, {\rm Hz}$ to get upper bounds on $r$.
We see that LMXBs spend about 6 times as much time radiating if their evolution is controlled
by thin rather than fat cycles.  However, current observations suggest that the mostly thinly clustered set of LMXBs
probably have  $\Omega_{\rm min} /2\pi \lesssim 250 \, {\rm Hz}$
and $\Omega_{\rm max} / 2\pi \gtrsim 350 \, {\rm Hz}$ \cite{vanderklis,strohmayer,bildsten}.
This case falls between the cases that Eqs.~(\ref{rthin}) and (\ref{rfat}) are valid. Instead, using these numbers
in Eqs.~(\ref{tspinup}) and (\ref{tspindown}) gives $r = 0.18$.
(It is possible of course that a subset of these might belong to an even more thinly clustered set of LMXBs; on the
other hand, some pulsars have $\Omega / 2\pi \gtrsim 600 \, {\rm Hz}$ indicating that some LMXBs
have been able to spin up well beyond $350 \, {\rm Hz}$.)
Levin \cite{levin} gives
an estimate of $10$--$100$ as the number of strongly accreting LMXBs in our galaxy.
This would indicate that if the $r$-modes instability controls the cycles of LMXBs
there is a good chance some LMXBs in the galaxy are currently radiating as long as the saturation
amplitude is small.
Levin \cite{levin} and Anderson, {\it et al.} \cite{ajks} come to more pessimistic conclusions, based on a larger saturation amplitude.
However, we should not be too optimistic either, and treat our values for $r$ as upper bounds.
Many LMXBs have accretion rates far lower
than $10^{-8} M_\odot / {\rm yr}$, lowering the real chances that any LMXB currently radiates.
Decreasing the size of the saturation amplitude increases the odds, but also makes the $r$-modes harder to detect.
We discuss detectability again at the very end of this section.  (Heyl \cite{heyl}, at the same time this paper was written,
independently considered small saturation amplitudes.  That paper estimates the number of LMXBs in the galaxy based
on the observed number of millisecond pulsars and comes to a similar but perhaps more optimistic conclusion than presented here.)

We now discuss the cycles in each figure. In Fig.~\ref{FigEvolveOrdBFlds} curves are shown for the ordinary-fluid case.
The dashed curves with large cycles are for $\dot{M} = 10^{-8} M_\odot \, {\rm yr}^{-1}$, $\alpha_{\rm sat} = 1$, and ${\cal S} = 1$.
The largest of these is for $B = 0 \, {\rm G}$ and the slightly smaller cycle is for $B = 10^{10} {\rm G}$.  The lower dashed curve
showing a thin cycle is for ${\cal S} = 0.1$ and $B = 10^{11} {\rm G}$.  This is the largest buried field that
could exist in the crust of an LMXB \cite{czb}, and we choose the slip factor to put this cycle within the observed
range of spin frequencies of LMXBs.  The thin solid lines are cycles for the same parameters as the dashed cycles, except for a more
realistic saturation amplitude of $\alpha_{\rm sat} = 10^{-4}$.  Recent studies indicate the saturation amplitude
is small, and perhaps much smaller than $10^{-4}$ \cite{wumatarras,arrasetal}.  However, when the saturation amplitude is this small,
the entire cycle moves into the superfluid region.  Since the transition temperature is uncertain we do not exclude these
curves as a possibility.  The main conclusion drawn here is that only a very large fields can produce thin cycles
in the ordinary-fluid case.

In Figs.~\ref{FigEvolveSupB00}-\ref{FigEvolveSupB10} results are shown respectively for the superfluid model with pinned neutron
vortices for $B = 0$, $10^{9}$, and $10^{10}$ Gauss, and for ${\cal S} = 0.01$, ${\cal S} = 0.005$, and ${\cal S} = 0.01$.
In each case, the slip factor is choose to put the
curves within the observed range of spin frequencies observed in LMXBs.  The case of pinned proton vortices would result in
similar plots, but for different values of the slip factor. The dashed curves correspond to
$\dot{M} = 10^{-8} M_\odot \, {\rm yr}^{-1}$ and $\alpha_{\rm sat} = 0.1$.  The thick solid curves are for the
same accretion rate, but $\alpha_{\rm sat} = 10^{-4}$.  The thin solid curves are for $\dot{M} = 10^{-11} M_\odot \, {\rm yr}^{-1}$
and $\alpha_{\rm sat} = 10^{-4}$.  The surprising main conclusion drawn here is that $B \sim 10^{9}$ has the largest effect
on narrowing the cycles.  This is because this value of $B$ tends to flatten the critical anglular velocity curve in the relevant
temperature range, while mutual friction causes it to vary with temperature again for larger $B$ fields.

In general, lowering the saturation amplitude moves the lower right-hand corner of cycle to the
left along the critical angular velocity curve, while increasing
the accretion rate moves the upper left-hand corner of a cycle to
the right along a critical angular velocity curve. Thus adjusting these can make a cycle
more or less narrow, as can the magnitude of the magnetic field.  This makes it hard to
use current observations of LMXBs to place contraints on the internal physics of these stars.
However, if a class of LMXBs with similar internal magnetic fields could be identified, one prediction
is that weakly accreting LMXBs should exhibit a wider range of spins than strongly accreting LMXBs.
In any case, magnetic fields can make the cycles thinner, and are thus potentially
important to the explanation of the observed clustering of LMXB spin frequencies.
Thus, if the appropriate magnetic fields are present in a class of LMXBs it
will tend to make the cycles thinner and increase the number of currently radiating sources.

Given the uncertainties we cannot predict with any certainty if any LMXBs are currently radiating.
However, no matter the odds, the detection of gravitational radiation from the $r$-modes is not ruled out
for gravitational-wave detectors currently coming online (with their planned enhancements)
as long as the saturation amplitude is not too small.
Using Eq.~(4.9) from Owen {\it et al.} \cite{owen-etal} but correcting for the fact that in our models
the $r$-modes are confined to the core, the average dimensionless gravitational-wave
amplitude for a source a distance $D$ away is,
\beqa
h = 7.6 \times 10^{-27} \left ( {\alpha_{\rm sat} \over 10^{-4} } \right )
\left ( {\Omega \over 600 \pi / {\rm s}} \right )^3
\left ( {10 \, {\rm Kpc} \over D } \right ). \label{gwhamp}
\eeqa
For this fudicial value of $h$, and assuming an integrated observation time of $10^7 \, {\rm s}$, the optimal signal-to-noise
(SNR) ratios are in the range $0.57-4.4$ for the LIGO noise curves given in Owen {\it et al.} \cite{owen-etal}. (Since it is not
possible to keep the detector optimally aligned with the source actual SNRs would be $5-10$ times smaller.
However, advanced detectors can greatly improve the SNR in narrow frequency bands.)

Thus, assume that gravitational radiation is detected from an LMXB in our galaxy.
The radiation can be identified as due to
the $r$-mode instability if the observed frequency is in the correct ratio to the spin frequency of the LMXB.
The measured gravitational-wave amplitude then determines the $r$-mode saturation amplitude
(if the distance to the LMXB is known) which determines the right side of a spin cycle.
If the accetion rate is known the left side of a spin cycle is known.
It is unlikely to observe an LMXB along the top or bottom of a spin cycle.
However, the maximum (or minimum) spin frequencies, which give the top (or bottom) of a spin cycle must
be larger( or smaller) than the observed values for these.
Thus, it might be possible to place limits on the interior magnetic fields of LMXBs
with known accretion rates depending on how well the allowed range of spin frequencies can be determined.

\section{Conclusions}

We have studied the $r$-modes in accreting neutron stars with magneto-viscous boundary layers for ordinary-fluid and superfluid models.
The ordinary-fluid model is discussed in detail in Mendell \cite{mendell2001}.
We have shown,
within our approximations,
that no solution to the MHD equations
exists when both the neutron and proton vortices are pinned.
However, solutions can exist for cases
when just one species of vortex is pinned.  In these cases, the instability of the $r$-modes to gravitational-wave emission can
limit the spins of neutron stars (though strongly pinned neutron vortices would completely suppress the instability).
If both the neutron and proton vortices are completely unpinned, the
results would be basically that found in Lindblom and Mendell \cite{lm2000} and Levin \cite{levin}.
In general, magnetic fields increase the dissipation rate and flatten the critical angular velocity
vs. temperature curves in both models.
However, mutual friction in the superfluid model tends to counteract the magnetic effects for high temperatures or high fields.
Furthermore, if the $r$-mode instability controls the spin cycles of LMXBs we have shown several things.
As in Levin \cite{levin} Anderson, {\it et al.} \cite{ajks}, and Heyl \cite{heyl}, we show that
decreasing the saturation amplitude greatly increases the odds that at least one LMXB in our galaxy is currently radiating gravitational waves.
We also show that making the spin cycles thinner further increases the odds by up to a factor of 6 compared to previous estimates.
Finally, we show for the first time that magnetic fields, by flattening the critical angular velocity curves,
make the spin cycles of LMXBs thinner and that this increases the fraction of time an LMXB spends radiating gravitational waves
for  $B \gtrsim 10^{11} \, {\rm G}$ in the ordinarly-fluid case and for $B \sim 10^9 \, {\rm G}$ in the superfluid case.
Previous studies predicated that probably no LMXBs are currently radiating in our galaxy, based on large values of the saturation
amplitude $\alpha_{\rm sat}$.
However, the subsequent studies of Wu, Matzner, and Arras \cite{wumatarras} and
Arras {\it et al.} \cite{arrasetal} predict that $\alpha$ will be small.
In any case, the previous studies agree with our result that
the fraction of currently radiating LMXBs approaches one if $\alpha_{\rm sat} \lesssim 10^{-4}$ and
if at least $10$ strongly accreting LMXBs exist in our galaxy.  However, if $\alpha_{\rm sat}$ is much larger than $10^{-4}$ then probably no
LMXBs are currently radiating; if $\alpha_{\rm sat}$ is much smaller than $10^{-4}$ the $r$-modes may be undetectable.
(Detection of a source in our galaxy is difficult, but not impossible for $\alpha$ as small as $10^{-4}$.)  However, if
gravitational waves are detected from the $r$-modes of an accreting neutron star, it might be
possible to place limits on the interior magnetic fields of LMXBs with known accretion rates
based on their observed allowed range of spin frequencies.

Of course many caveats must be added to our conclusions.  Our evolution graphs show that the fluids may go through a transition
from superfluid to ordinary-fluid during the heating phase, and vice versa during the cooling phase.
Vortices would probably not have time to
migrate out of the core during the spindown phase, and thus superfluid effects could remain important above
the transition temperature; but the details of what happens to the boundary layer during such a phase transition could become very messy.
This is only one complication that we have ignored.  The neutrons could be in the superfluid phase
while the protons are in the normal phase (for example see Easson and Pethick \cite{eassonpethick}) and vice versa.
Another possibility is that the protons could form
a type I superconductor in the intermediate state rather than a type II superconductor in the vortex state \cite{bppNature69}.
We have also only studied the effects of mutual friction for one value of the entrainment factor; special values exist that would
completely change our results \cite{lm2000}.
Furthermore, we have ignored interactions between neutron and proton vortices (see \cite{rudermanetal97})
and other dissipative effects, such as hyperon bulk viscosity \cite{jones2001a,jones2001b,lohyperons,haenseletalhyp}.
Finally, nonlinear effects are ignored, such as the winding of magnetic fields lines, which could be important especially
for $B \gtrsim 10^{10} \, {\rm G}$ \cite{spruit,rls,rmsmag}.  The winding of field lines would introduce another temperature independent
damping mechanism, and thus probably would complement our results. Clearly, an exact understanding of the $r$-mode instability is very complicated.

Finally, we remark on one final interesting question:
If the $r$-modes control the spins of LMXBs, could the observation of gravitational waves
from some of these objects distinguish between ordinary-fluid neutron stars, superfluid
neutron stars, and strange stars?  The recent study of Andersson, Jones, and Kokkotas \cite{ajkstrange}
suggests that strange stars in LMXBs may emit persistent gravitational waves from saturated $r$-modes.
We have shown that neutron stars in LMXBs, in agreement with Levin \cite{levin}, Anderson, {\it et al.} \cite{ajks},
and Heyl \cite{heyl}, would emit gravitational waves for only part of their spin cycle.  If we do not observe gravitational
radiation from any LMXBs we probably do not learn much, since so many mechanisms can suppress the instability.
However, if the $r$-modes are observed from some LMXBs, and the current theoretical understanding holds up,
then the fraction of LMXBs that radiate would distinguish between neutron and strange stars.
A detailed comparison between theory and observation would be needed to distinguish between ordinary-fluid
and superfluid neutron stars, however. Obviously, further work is needed to understand the various possibilities.

\label{sectionVI}

\acknowledgments We wish to thank B.~Owen, L.~Lindblom, R.~Wagoner, and R.~Epstein for responding to email questions and
helpful discussions concerning this work.  JBK thanks the NSF REU program and Caltech for financial support and
LIGO Hanford Observatory for hospitality extended to him while completing this work.
This research was supported by NSF grant PHY-0096304.

\appendix

\section*{ Superfluid MVBL Solution}

%-----------------------------------
We will now solve the corrective differential equations governing
MVBL dissipation in a superfluid neutron star with arbitrary $B$-field and without our
ignoring a factor of ${\rm cos}^2 \theta$ in Eq.~(\ref{Wthetaeqn}).
For such a model, we obtain the system of differential equations:
\begin{eqnarray*}
A \delta \tilde{v}^{\theta} + B \delta \tilde{v}^{\phi} &=&
 F \delta \ddot{\tilde{v}}^{\theta} + G \delta \ddot{\tilde{v}}^{\phi} +
H \delta \ddot{\tilde{w}}^{\theta} + I \delta \ddot{\tilde{w}}^{\phi}  \\
-B \delta \tilde{v}^{\theta} + A \delta \tilde{v}^{\phi} &=&
J \delta \ddot{\tilde{v}}^{\theta} + K \delta \ddot{\tilde{v}}^{\phi} +
L \delta \ddot{\tilde{w}}^{\theta} + M \delta \ddot{\tilde{w}}^{\phi}  \\
C \delta \tilde{v}^{\theta} + D \delta \tilde{v}^{\phi} &=&
F \delta \ddot{\tilde{v}}^{\theta} + G \delta \ddot{\tilde{v}}^{\phi} +
H \delta \ddot{\tilde{w}}^{\theta} + I \delta \ddot{\tilde{w}}^{\phi}  \\
-D \delta \tilde{v}^{\theta} + E \delta \tilde{v}^{\phi} &=&
J \delta \ddot{\tilde{v}}^{\theta} + K \delta \ddot{\tilde{v}}^{\phi} +
L \delta \ddot{\tilde{w}}^{\theta} + M \delta \ddot{\tilde{w}}^{\phi}
\end{eqnarray*}
where
\begin{eqnarray*}
A &=& i\rho\kappa  \\
B &=& -2\rho\cos\theta \\
C &=& i\rho_p\kappa + 2\gamma\rho B_n\cos^2\theta \\
D &=& -2\rho_p\gamma\cos\theta \\
E &=& i\rho_p\kappa + 2\gamma\rho B_n \\
F &=& \frac{1}{\Omega} \left\{ \frac{-i\rho_p V_{CV}^2}{\kappa\Omega} \left(\frac{B^r}{B}\right)^2 \left[1 - \frac{(B^{\theta})^2}{B^2}\right] + \eta_e \right\} \\
G &=& \frac{i\rho_p V_{CV}^2}{\kappa\Omega^2} \left[ \frac{(B^r)^2 B^{\theta} B^{\phi}}{B^4} \right] \\
H &=& \frac{\rho_n}{\rho}\gamma F \\
I &=& \frac{\rho_n}{\rho}\gamma G \\
J &=& G \\
K &=& \frac{1}{\Omega} \left\{ \frac{-i\rho_p V_{CV}^2}{\kappa\Omega} \left(\frac{B^r}{B}\right)^2 \left[1 - \frac{(B^{\phi})^2}{B^2}\right] + \eta_e \right\} \\
L &=& \frac{\rho_n}{\rho}\gamma J \\
M &=& \frac{\rho_n}{\rho}\gamma K.
\end{eqnarray*}
As in the simplified case, this system of differential equations only has two linearly independent variables:
\begin{eqnarray*}
\delta\tilde{v}^{\theta} &=& a \delta\tilde{w}^{\theta} + b \delta\tilde{w}^{\phi} \\
\delta\tilde{v}^{\phi} &=& c \delta\tilde{w}^{\theta} + d \delta\tilde{w}^{\phi}
\end{eqnarray*}
where
\begin{eqnarray*}
a &=& \frac{AC + BD}{A^2 + B^2} \\
b &=& \frac{AD - BE}{A^2 + B^2} \\
c &=& \frac{BC - AD}{A^2 + B^2} \\
d &=& \frac{AE + BD}{A^2 + B^2}.
\end{eqnarray*}

This leaves us with two linearly independent systems of differential equations.
Assuming the same exponential behavior for $\delta\tilde{v}$ and $\delta\tilde{w}$
as in the simplified model, we find
\[ k_{\pm}^2 = \frac{\beta \pm \sqrt{\beta^2 - 4\alpha\gamma}}{2\alpha} \]
where
\begin{eqnarray*}
\alpha &=& PS - QR \\
\beta &=& PE + SC + DQ - DR \\
\gamma &=& D^2 + CE
\end{eqnarray*}
and
\begin{eqnarray*}
P &=& aF + cG + H \\
G &=& bF + dG + I \\
R &=& aJ + cK + L \\
S &=& bJ + dK + M.
\end{eqnarray*}

%%%%%%%%%%%%%%%%%%%%%%%%%%%%%%%%%%%%%%%%%%%%%%%%%%%%%%%%%%%%%%%%%%%%%%%%%%%%
%%


\begin{references}
\bibitem{cycleshape} The code that generated the cycles used quality step-size control during the heating and spindown phases of the evolution.
To speed up the code, once the cycle turned the lower right corner and the cooling phase had begun (indicated by the angular velocity reaching its
minimum value and not changing) the code extrapolated to get the cooling and spinup phase of the curves.
This was found to give the same cycles as using quality step-size control thoughout the evolution, except it makes
the lower left corner of the cycles turn too sharply.
\end{references}
\end{document}